\documentclass[11pt]{article}

\usepackage[footnotesize,it]{caption}
\usepackage[pdftex]{graphicx}      
\usepackage[pdftex]{color}         
\usepackage{amsmath}              
\usepackage{amssymb}              
\usepackage{amsfonts}             
\usepackage{verbatim}             
\usepackage{enumerate}
\usepackage{natbib}
\usepackage{setspace}
\usepackage{subfig}
\usepackage{epstopdf}
\usepackage{hyperref}
\usepackage{authblk}

\begin{document}



\title{A network-patch methodology for adapting agent-based models for directly transmitted disease to mosquito-borne disease}

\author{Carrie A. Manore$^{1}$$^{\ast}$\thanks{$^{\ast}$Corresponding author Email: cmanore@tulane.edu\vspace{3pt}} 
, Kyle S. Hickmann$^{1}$$^{\diamond}$\thanks{$^{\diamond}$Corresponding author Email: khickma@tulane.edu\vspace{3pt}}, 
James M. Hyman$^{1}$, Ivo M. Foppa$^{2}$, \\
Justin K. Davis$^{1}$, Dawn M. Wesson$^{3}$, and Christopher N. Mores$^{4}$ \\
\vspace{6pt} 
$^{1}${\em{Center for Computational Science, Department of Mathematics, Tulane University}};
$^{2}${\em{Battelle/Epidemiology \& Prevention Branch, Influenza Division, CDC}};
$^{3}${\em{Department of Tropical Medicine, School of Public Health and Tropical Medicine, Tulane University}};
$^{4}${\em{Vector-borne Disease Laboratories, Center for Experimental
  Infectious Disease Research, Department of Pathobiological Sciences,
  School of Veterinary Medicine, Louisiana State University}}} 

\maketitle

\begin{abstract}
  Mosquito-borne diseases cause significant public health burden, mostly in tropical and sub-tropical parts of the
  world, and are widely emerging or re-emerging in areas where previously absent. Understanding, predicting, and
  mitigating the spread of mosquito-borne disease in diverse populations and geographies are ongoing modeling
  challenges. We propose a hybrid network-patch model for the spread of mosquito-borne pathogens that accounts for the
  movement of individuals through mosquito habitats and responds to environmental factors such as rainfall and
  temperature.  Our approach extends the capabilities of existing agent-based models for individual movement developed
  to predict the spread of directly transmitted pathogens in populations.  To extend to mosquito-borne disease, these
  agent-based models are coupled with differential equations representing `clouds' of mosquitoes in geographic patches
  that account for mosquito ecology, including heterogeneity in mosquito density, mosquito emergence rates, and the
  extrinsic incubation period of the pathogen. We illustrate the method by adapting an agent-based model for human
  movement across a network to mosquito-borne disease.  We investigated the importance of heterogeneity in mosquito 
  population dynamics and host movement on pathogen
	transmission, motivating the utility of detailed models of individual behavior and observed that the random mixing 
	model only captured the dynamics of the the high movement rate scenario. We observed that  the total number of 
	infected people is greater in heterogeneous patch models 
 	with one high risk patch and high or medium human movement than it would be in a random mixing homogeneous model.  
	Our hybrid agent-based/differential equation model is able to quantify the importance of the 
	heterogeneity in predicting the spread and invasion of mosquito-borne pathogens.  Mitigation strategies can be more 
	effective when guided by realistic models created by extending the capabilities of existing agent-based models to include vector-borne
	diseases.

%
\end{abstract}

\section{Introduction}

Predicting the spread of mosquito-borne pathogens such as dengue virus, chikungunya virus, Rift Valley fever, and West
Nile virus is complicated by complexity of the systems, lack of appropriately granular data, and computational expense
of realistic models. The existing models for the spatial spread of mosquito-borne pathogens, while providing important
insight into disease dynamics, often ignore either detailed host movement and/or explicit mosquito population dynamics
to reduce complexity and computational expense \cite{adams2009man,anyamba2009prediction,vanwambeke2011spatially}. These
and other studies have shown that capturing host behavior and movement through the mosquito environment is important and
perhaps even crucial to understanding risk and informing mitigation efforts
\cite{adams2009man,ReinerJr2014,perkins2013heterogeneity,stoddard2013house}. We bridge this gap by combining an
agent-based spatial model (ABM) for host movement on a network with a patch based ordinary differential equation (ODE)
model that captures environmental and mosquito dynamics in geographic habitat patches that cover the region modeled by
the ABM. In particular, we introduce a relatively simple methodology for extending already existing large-scale ABMs for
hosts to include mosquito-borne disease.  This `network-patch' model will help quantify the importance of heterogeneity
in these components and aid in predicting the spread of mosquito-borne pathogens, particularly in forecasting the
initial spread following introduction into a new location. The hybrid model described here focuses on the transmission
dynamics of mosquito-borne pathogens, but extensions to other arthropod vectors and zoonotic or animal vector-borne
pathogens is possible.

The network-patch methodology can be used to adapt existing ABMs for person-to-person transmitted diseases, such as
influenza (e.g. EpiSimS \cite{eubank2004modelling,mniszewski2008pandemic}, FRED \cite{grefenstette2013fred,FRED}, DISimS
\cite{bissetinteractive}), to mosquito-borne pathogens. Similarly, ABMs for animal movement could be adapted to include
mosquito-borne pathogen spread in addition to directly transmitted pathogens. The method accounts for explicit spatial
arrangement of mosquito habitat, the social aspects of host behavior, and variations in environment and weather.  Adams
and Kapan \cite{adams2009man} modeled spatial mosquito-borne disease on a network where each network node corresponded
to exactly one patch and where the mosquito populations did not explicitly depend on weather or landscape.  We expand
and extend their model for mosquito patches that incorporate variation in mosquito density, determined by biotic
(e.g. vegetation, predators) and abiotic factors (e.g. temperature, humidity, breeding sites)
\cite{anyamba2009prediction,vanwambeke2011spatially}.  Perkins et al. \cite{perkins2013heterogeneity} explored the idea
of different habitat patches for various mosquito life cycle stages (blood seeking, resting, oviposition) with
aggregated movement of humans between patches based on proportion of time spent in each of the patches. We extend this
idea by coupling mosquito habitat patches with ABMs that have already been tuned to model human behavior and movement in
a city or region.

There are several ABMs for mosquito-borne disease where mosquitoes are treated as independent agents
\cite{almeida2010multi,chao2012controlling,cummins2012spatial,eckhoff2011malaria,padmanabha2012interactive}. The results
of such models highlight the role of heterogeneity in host movement, mosquito distribution and density, and the
environment in mosquito-borne disease spread.  However, they are restricted in spatial scale by the computational cost
of modeling each individual mosquito and host. Since relatively little data is available for individual mosquito
host-seeking behavior across larger scales, there is significant parametric uncertainty associated with models for
individual mosquitoes, particularly on a large scale.  Our approach can more easily incorporate larger host populations
and a wider geographical area. We account for heterogeneity in disease spread on the spatial scale of patches, rather
than the exact spatial location of the individual mosquitoes, as in the related research described in
\cite{arino2003multi,hyman2003modeling,manore2011disease,mcmahon2013,xue2012network}. Our model presents a unified
framework useful for simulating emerging epidemics, and understanding the roles of spatial, ecological, environmental,
social, and behavioral heterogeneities in mosquito-borne diseases.

\begin{figure}[h]
  \begin{center}
    \includegraphics[scale=0.5]{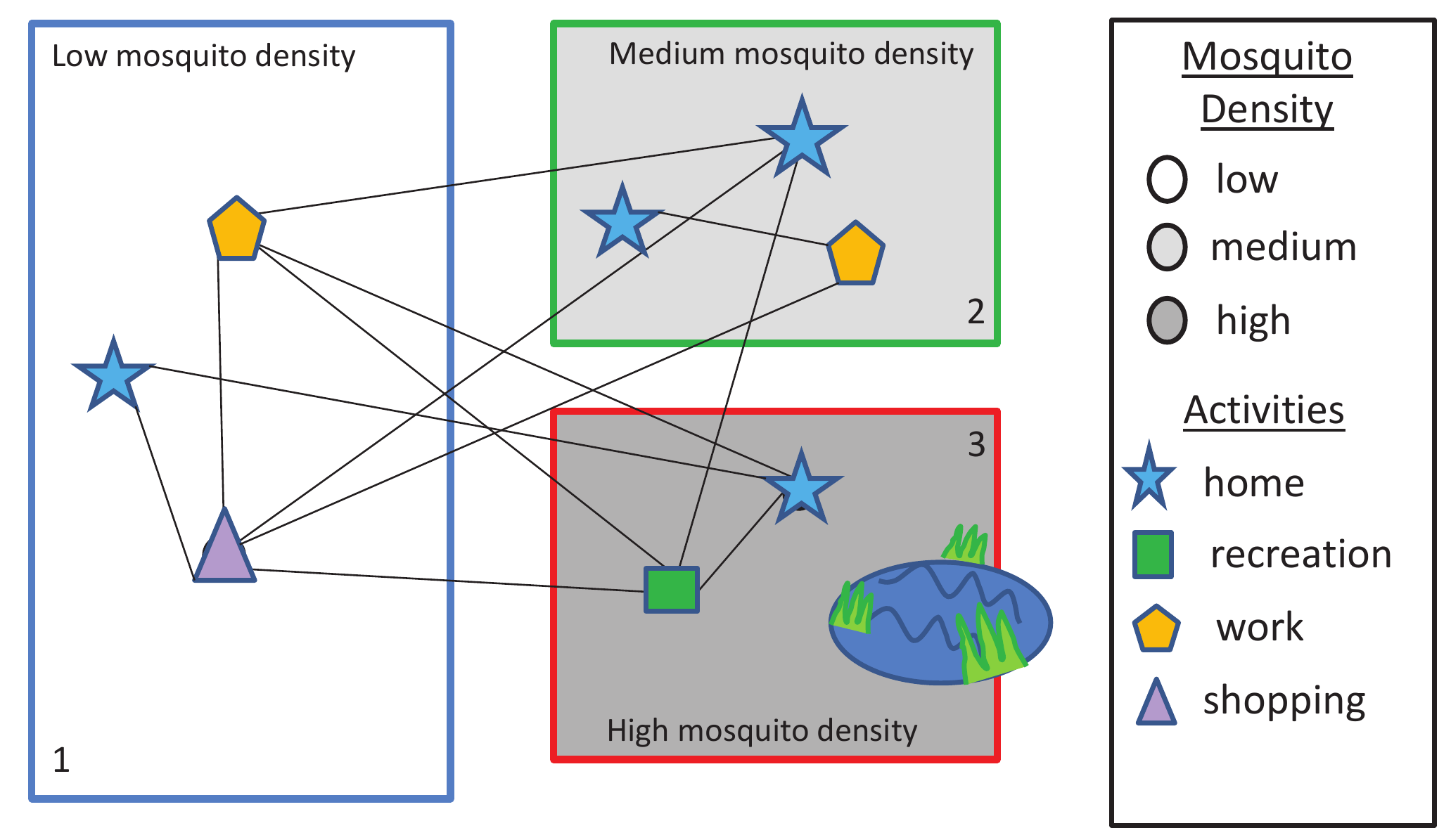}
  \end{center}
  \caption{ The network-patch model combines the detailed host movement captured by an agent-based spatial network model
    with a habitat patch model for mosquitoes.  The agents in the network model move between locations and activities
    (network nodes) determined by population, demographics, and host behavior.  We give examples of human activities
    here. Animal activities could include foraging, drinking, and sleeping locations.  Each node is associated with an
    environmental patch where the local population of infected and uninfected mosquitoes determine the risk of an
    individual becoming infected while in the patch.  }
  \label{F:NetworkPatch}
\end{figure}

To create the network-patch model, we overlay geographic patches based on the environmental properties correlated with
the mosquito's life cycle on a connected host network.  The location and size of the mosquito patches is determined by
landscape, vegetation, weather, human socioeconomic factors, land use, and availability of mosquito breeding sites. The
level of detail used in modeling the mosquito patches will depend on available data and expert opinion.  Many of the
existing agent-based models have the agents located at nodes based on location (e.g., office building, school, home).
For the network-patch model, these location nodes are assigned to a specific patch, and the activity of the individual
is then mapped to the appropriate activity category in the patch. Each activity category within a patch is assigned a
relative risk of being bitten by a mosquito.  Figure \ref{F:NetworkPatch} illustrates how each patch can encompass
several network nodes and edges so that any individual at or between nodes is exposed to the hazards in the patch.

The mosquito dynamics in each patch are simulated using systems of ODEs, as opposed to using an individual-based model
representation for each mosquito.  The ODE \emph{mean field} model is an upscaling, or homogenization, of the
computationally more expensive models that track many individual mosquitoes. In effect, the model is a summary
representation of the finer scale mosquito population dynamics, for which we often lack appropriately scaled demographic
data necessary to model individual mosquitoes.  The mosquito population in a patch depends upon the local environment,
weather, and mitigation strategies, thus, the ODEs are based on the assumption that the mosquito dynamics take place on
a much smaller scale than the individual dynamics of the host agents. The risk of being bitten by a mosquito depends
upon the local mosquito density, host density, and type of host activities. The probability of disease transmission
depends on the disease prevalence in the patch, the biting rate of the mosquitoes, and the susceptibility of the
individuals.

We used the method described here to adapt an ABM for host movement on a network to vector-borne diseases. Simulations
illustrate scenarios for which including the heterogeneity of an ABM is important to understanding the risk of an
outbreak, disease dynamics, and effective mitigation strategies. We show that with spatial heterogeneity in host and
mosquito density, varying host movement rates between patches can produce different results than a standard ODE model or
ODE multi-patch model.  We are currently using this method to adapt a complex ABM, EpiSimS
\cite{mniszewski2008pandemic}, to model mosquito-borne disease in the United States. It will also be important to
develop techniques for analysis of this hybrid model framework, including uncertainty quantification, sensitivity
analysis, and determining the basic and effective reproduction numbers. Once implemented, this method will expand the
capabilities of ABMs for complex host movement and decision making to include an important class of diseases.

After developing the overall structure of the network-patch model, we describe how the force of infection is computed
and give an approximate formulation for the basic reproduction number for each patch. We then present a detailed example
of the methodology and results from simulations that illustrate the importance of accounting for host movement and
heterogeneity in mosquito habitat for understanding and controlling mosquito-borne disease.

\section{The Model}

The patch and agent-based models are propagated simultaneously and coupled through the mosquito biting rate and disease
transmission.  After describing the overall framework of the ABM host network and ODE mosquito models, we provide a
detailed description of how the infection spreads between the two populations.  In the next section, we present a simple
ABM to illustrate the governing principles of the network-patch model.

\subsection{The Agent-Based Host Model}

We follow the general idea of the ODD protocol outlined in \cite{grimm2010odd} to describe the agent-based (or
individual-based) portion of the model. Host movement is defined by an ABM on a network of locations and the activity of
each agent is tracked during their daily activities. The exact implementation of host movement will depend upon the ABM
being used and on the questions asked. In general, however, the location of each host agent is updated at chosen
discrete time intervals based on set movement and activity rules
\cite{bissetinteractive,eubank2004modelling,mniszewski2008pandemic,FRED,grefenstette2013fred}.  At any specific time,
each agent or individual exhibits characteristics, such as their current activity, susceptibility to infection,
infection status, and patch location, $k$. The agents move between connected location or activity nodes as determined by
an underlying movement model.  The activity patterns might come from a complex agent-based simulation, such as EpiSimS,
or a simple random walk algorithm where the individuals randomly change their location/activity at the end of each
time-step. This method is designed to be flexible enough to use for different ABMs.

\subsubsection{Purpose, entities, state variables and scales of ABM}

For this paper, we use an ABM developed for host movement on a network where each node is a location. The purpose of the
original ABM was to simulate human movement on a network in order to model directly transmitted infectious diseases such
as influenza across various network connectivity and host movement regimes. The model depicts a certain number of agents
representative of humans moving between a network of locations while spreading an infectious disease.

Each agent is assigned an initial node location. At each time step, individuals can either move or not move depending on
the user defined probability of movement. If the individual does decide to move, a node to move to is chosen. The model
assumes that agents can only move directly to nodes with an edge connecting to the currently occupied node.  If an
infected individual is at a node the mosquito population in the patch that the node belongs to has an increased rate of
growth in the infected population of mosquitoes. This in turn creates increased risk of infection for individuals based
at nodes within the same mosquito patch.  Once infected, an individual will progress to an exposed/incubating stage
where they are infected but not yet infectious. Next, the exposed/incubating agent will move to the infectious (and
usually symptomatic) stage and, finally, to a recovered/immune stage. The distribution of times spent in each of these
stages is user defined. The model can be tuned to any desired time scale. For this paper, we have chosen a time scale of
days with the model updating every 6 hours ($0.25$ days).

\subsubsection{Process overview and scheduling}

To set up a run, the chosen number of locations/nodes are randomly connected with probability $p$ using the
Erd\H{o}s-R\'enyi graph algorithm from NetworkX \cite{hagberg2004networkx}. Initially, the agents are randomly assigned
a movement rate (the average rate at which a person moves to a neighboring location), a current disease state (set equal
to susceptible for all but a few randomly selected infected individuals), and a randomly chosen initial location all
chosen from user-defined distribution.  Here, the movement, incubation (time in the exposed state), and recovery (time
in the infectious state) rates are assumed to have a lognormal distribution (Table \ref{T:PatchParams}).

The agents are advanced in fixed increments, $\Delta t$, in time.  During a time-step, the infection status of each
individual is updated.  Human movement within the model takes place at the end of the time-step after the disease states
have been updated.  The movement rate $\rho$ is chosen from a probability distribution $M$ that determines the
likelihood that an agent will change location at each time step.  That is, an agent will moved to another node with
probability $Pr = 1 - e^{- \Delta t \rho}$, where $\rho$ is a random sample from the probability distribution given in
Table \ref{T:PatchParams}.  Thus, a higher movement rate indicates that an agent is more likely to move to another
node. If an agent does move, then the node to which it moves is selected with uniform probability from its neighbors.

\subsection{The Mosquito Model}
 
The location and size of mosquito patches overlayed on the ABM will be determined for each mosquito species. Although
this can be implemented at any scale, we anticipate patches on the order of a building or a group of buildings
(e.g. city block) when modeling a city or local habitat patches for animals. As hosts enter and spend time in a patch,
their infection risk depends on the probability of being bitten by an infectious mosquito and their individual
susceptibility to infection.  The number of bites an individual gets depends upon the activity they are engaged in, the
number of mosquitoes and other hosts in the patch, and other environmental factors, such as the temperature or time of
day.  Activities (or locations) are mapped to a relative mosquito exposure parameter, $\alpha \in [0,1]$, that defines
the relative risk of an agent being bitten by a mosquito given an underlying risk for mosquito bites in the patch and
the agent's activity. For example, for humans, outdoor activities might have a much higher biting rate than an indoor
activity, especially if the building has screens and air conditioning. In general, risk will also depend on the mosquito
species and habitat being considered.

We assume that the mosquito population dynamics depend primarily on the mosquito species, temperature, photoperiod, and
rainfall.  Rainfall, or for some species, paradoxically, a lack of rainfall, can lead to mosquito population increases,
which can lead to increased biting intensities experienced by humans. Mosquitoes successfully feeding on blood can lead
to more eggs being oviposited and an increase in immature stages of mosquitoes
\cite{gong2167climate,shaman2007reproductive}.  Temperature affects the mosquito maturation rate (egg, larva, pupae),
the extrinsic incubation period \cite{ba2005aspects,turell1985effect}, and the mosquito life span
\cite{depinay2004simulation,kasari2008evaluation}.  Vegetation, sanitation, land use, and building density also
influence the mosquito life cycle.  Some mosquito species, including \textit{Aedes aegypti}, can complete most or all of
their life cycle inside human dwelling, resulting in less dependence on weather and more dependence on human behavior.
Field studies, detailed ABMs for \textit{Aedes aegypti} population dynamics (e.g., Skeeter Buster
\cite{magori2009skeeter}), and differential equation mosquito-borne disease models help define the relevant ODE model
parameters.

The adult female mosquito population is divided into three epidemiological classes, susceptible ($S$), exposed ($E$),
and infectious ($I$).  The female mosquitoes emerge into the susceptible class, $S_v$, (where the subscript v stands for
vector or mosquito) and are exposed to infection after biting an infected human. Mosquitoes that are infected with the
pathogen, but are not yet able to infect a susceptible human host by biting are in the exposed/incubating class. This
period is referred to as the extrinsic incubation period (EIP).  Once infected, a mosquito moves from incubating to
infectious at the rate $\nu_v$ with average time spent in the incubating, $E_v$, class being $1/\nu_v$. The EIP can
depend on temperature, strain of the pathogen, or mosquito species \cite{turell1985effect}. After the incubation period,
the mosquito moves from the exposed class to the infectious class, $I_v$. The mosquito remains infectious for life, with
an average lifespan depending on species and environmental conditions.
 
Adult female mosquitoes die at a per capita rate $\mu_v$, where $1/\mu_v$ is the average lifespan of an adult female
mosquito in the given patch. This death rate can depend on exogenous factors such temperature, food availability, and
humidity. Mitigation strategies focusing on increasing adult death (adulticides) can be included here via reduction of
the adult female mosquito lifespan or chosen mosquito carrying capacity.  The total adult female mosquito population is
represented by $N_{v}=S_{v}+E_{v}+R_{v}$.

Each mosquito patch is characterized by the scalar parameters (Table \ref{T:param}) specific to that patch.  Patches are
chosen so that the mosquito dynamics can be approximated as being homogeneously mixed and uniformly spatially
distributed within a patch.  Each model parameter, and thus the mosquito dynamics output, can depend upon the associated
habitat patch, denoted by a patch index $k$. For the full model, every variable and parameter will have a superscript
$k$ to denote the particular patch location. However, for simplicity of notation, we suppress the superscript here.  Our
mosquito dynamics model for a patch is described as a system of ODEs as follows:
\begin{subequations} \label{E:mosquito2}
 \begin{align}
\frac{d S_{v}}{dt} &= h_v(N_v,t) - \lambda_v(t) S_{v} - \mu_{v} S_{v} \\
\frac{d E_{v}}{dt} &=  \lambda_v(t) S_{v} -\nu_vE_v  - \mu_{v} E_{v}\\
\frac{d I_{v}}{dt} &= \nu_v E_{v} - \mu_v I_{v}.
\end{align}
\end{subequations}
The total number of adult female mosquitoes, $N_v=S_v + E_v + I_v$, includes all mosquitoes in the patch.  The average
force of infection to mosquitoes, $\lambda_v(t)$ (rate of infection for each mosquito per unit time), in the patch is
defined as the product of the average number of bites per mosquito, the probability that a bite is on an infectious
host, and the probability of transmission per bite.  The details of defining $\lambda_v(t)$ will be given in the next
sections.

The adult female mosquito per-capita emergence function, $h_v(N_v,t)$, can be constant or vary with time as a function
of the weather, host availability, density dependence in larvae, and other factors.  We define $h_v(N_v,t) =
\left(\psi_v - \frac{r_v N_v}{K_v}\right)N_v$
where $\psi_v$ is the natural per-capita emergence rate of female mosquitoes in the absence of density dependence. This
term depends on egg-laying rates, probability of hatching, surviving larvae and pupae stages, and successful emergence
as adult females.  $K_v$ is the carrying capacity of the mosquitoes in the patch and $r_v=\psi_v-\mu_v$ is the intrinsic
growth rate of mosquitoes in the absence of density dependence. We include density dependence in the emergence function
$h_v$, rather than the adult mosquito death rate, because density dependence has been shown to occur in the aquatic
larvae stage. However, the emergence rate can be simplified to be density independent, if desired, or adapted to a
particular mosquito species or situation as needed.  The complexity and structure of the mosquito model chosen depends
upon the system considered and questions being asked.

Summing the equations \ref{E:mosquito2}, and using the definition of $h_v(N_v,t)$, the total mosquito population in each
patch is modeled by:
\begin{equation}
\frac{dN_v}{dt} = \left(\psi_v - \frac{r_v N_v}{K_v}\right)N_v - \mu_vN_v = r_v\left(1-\frac{N_v}{K_v}\right)N_v
\end{equation}

In addition to varying between patches, all parameters can depend on time, weather, mosquito species, or other
factors. For example, one can incorporate a mosquito lifespan dependent on temperature, so that the mosquito per-capita
death rate $\mu_v=\mu_v(T)$ depends on temperature $T$. This model can easily be adapted to include additional classes
such as separate egg and larvae classes or several infectious classes for the mosquitoes. For some pathogens, vertical
transmission in mosquitoes can be important. This model can also be adapted to include vertical transmission (see,
e.g. \cite{chitnis2013modelling} for Rift Valley fever). Vector control measures can be incorporated explicitly to the
different life stages of the mosquito via the emergence rate, death rate, changine the number of mosquitoes, or patch
carrying capacity. Some of the more important time dependent parameter variations are seasonal mosquito recruitment
rate, seasonal biting rate, seasonal mortality rate, and a temperature-dependent seasonal EIP.

Our current model is deterministic, but stochastic variations in the environmental variables, mosquito populations, and
infectious status can all be important, especially in the early stages of an emerging epidemic when there are few
infected mosquitoes in a patch.  In these situations, the model can be modified to include these effects using
approaches such as the one described in Allen et al. \cite{allen2008introduction} for stochastic disease models.

\begin{table}
\caption{Parameters for the mosquito patch model and their dimensions. The range of parameter values and references
    are given in a table with the numerical simulations.}
{\begin{tabular}{lp{11.5cm}}
  $\psi_{v}$: & Per capita emergence rate of adult female mosquitoes.
  (Time$^{-1}$)\\
  $\mu_{v}$: & Per capita death rate of adult female mosquitoes.
  (Time$^{-1}$)\\
    $K_{v}$: & Maximum number of mosquitoes in the patch.
  (Mosquitoes)\\
  $\sigma_{v}$: & Number of times one mosquito would want to bite a host per unit time, if hosts were freely available. This is a function of the mosquito's gonotrophic
  cycle (the amount of time a mosquito requires to produce eggs). (Time$^{-1}$)\\
  $\sigma_h$: & The maximum number of mosquito bites an average host can sustain per unit time. This is a function of the host's exposed surface area, the efforts it takes to prevent mosquito bites, and any vector control interventions in place to kill mosquitoes or prevent bites. (Time$^{-1}$) \\
  $\beta_{hv}$: & Probability of transmission of infection from an infectious mosquito to a susceptible host given that a contact between the two occurs. (Dimensionless)
  \\
  $\beta_{vh}$: & Probability of transmission of infection from an infectious host to a susceptible mosquito given that a contact between the two occurs. (Dimensionless)
  \\
  $\nu_v$: & Per capita rate of progression of mosquitoes from the exposed state to the infectious state. $1/ \nu_v$ is the average duration of the latent period.
  (Time$^{-1}$) 
\end{tabular}}
\label{T:param}
\end{table}

\subsection{Biting and Infection Rates}

The ABM and mosquito ODE models communicate with each other through the force of infection parameters.  In each patch
$k$, the force of infection from mosquitoes to hosts, $\lambda_{h,j}^k$ (where the subscript $h$ refers to hosts and the
subscript $j$ refers to activity), is communicated to the ABM.  This is used to determine the actual probability of
being bitten by an infectious mosquito, based on each agent's activity and associated relative mosquito exposure
modifier, $\alpha_j$.  In the same way, the probability of a susceptible mosquito biting an infectious host in the patch
is used to compute the mosquito force of infection, $\lambda^k_{v,j}(t)$, and is communicated back to the ODE equations
(\ref{E:mosquito2}) via the total number of agents, $N_h^k$, and the number of infectious agents, $I_h^k$, in each patch
as well as the relative mosquito exposure of each agent, $\alpha_j$.
 
\subsubsection{Biting Rate}

The force of infection for the spread of the epidemic depends on the mosquito-host contact rate, or biting rate.  Each
location and activity in a patch will have an associated level of relative potential exposure to mosquitoes, $\alpha_j$
with $0 \le \alpha_j \le 1$. For example, if a location is outside, the potential for exposure to biting mosquitoes is
high so we assume $\alpha=1.0$. For humans, a more moderate risk location could be a building with no screens, faulty
screens, or no AC for which we may assume $0.5 \le \alpha \le 1.0$. Buildings with screens and/or AC are considered low
risk so $0.0 \le \alpha \le 0.5$ is assumed. The assumptions will depend upon host species, mosquito species, and
habitat. For example, mosquito populations which are ovipositing and hatching indoors may require different methods to
calculate risk of exposure.

We define $\sigma_v$ as the total number of (successful) bites a single mosquito would like to have, per unit time.
This can depend upon the gonotrophic cycle length (i.e., process of feeding on blood, resting, and egg laying), the
weather, and mosquito species.  The number of bites a host can sustain over a given time, $\sigma_h$, depends on exposed
skin area, attempts to deter biting (such as swatting and repellent), location (outside, inside with screens, inside
with AC), and other mitigation strategies.

In a patch $k$, $\sigma_v^k N^k_v$ is the total number of bites that all the the mosquitoes in a patch would like to
make per unit time and $\sigma_h^k \hat N^k_h$ is the maximum number of bites available to the mosquitoes. The variable
$\hat N^k_h$ represents the total number of hosts in patch $k$, scaled by their availability to be bitten.  If all the
hosts are at full risk of being bitten ($\alpha_j=1$), then $\hat N^k_h = N^k_h$, the total number of hosts in patch
$k$.  If there are people engaged in multiple activities in a patch with different risks of being bitten, then $\hat
N^k_h = \sum_j \alpha_j N^k_{h,j}$.  Here $N^k_{h,j}$ is the number of people in patch $k$ engaged in activity $j$.

We estimate the total number of contacts (successful bites) between hosts and mosquitoes in patch $k$ with the function
\cite{chitnis2007bifurcation,chitnis2008determining,chitnis2013modelling,manore2013RVF} as
\begin{equation} \label{E:totalmb}
b^k = b(N^k_v,\hat N^k_h) = \frac{\sigma_v N^k_v \sigma_h \hat N^k_h}{\sigma_v N^k_v +\sigma_h \hat N^k_h} 
~~.
\end{equation}
This formula has the correct limiting behavior as the number of mosquitoes or hosts approaches zero or infinity and has
meaningful contact rates for any vector-to-host ratio.  Other commonly used contact rates, such as frequency-dependent
contact or density-dependent contact, do not have the correct limiting behavior as populations vary greatly in time.
However, if the vector-to-host ratios are known to stay within a certain range, then any of these contact formulations
can be tuned to give approximately the same biting rates and $b^k$ can be changed to the desired contact formulation.

We define $b^k_v = b_v(N^k_v,\hat N^k_h) = b^k/N^k_v$ as the number of bites per mosquito per unit time in patch $k$.
As the mosquito population gets low or the host population gets very large (i.e. the vector to host ratio is small), the
number of bites is limited primarily by mosquito density. In this situation, the number of bites per mosquito is close
to $\sigma_v^k$ and the number of bites per host is close to $\sigma_v^k N^k_v/\hat N^k_h$. When the mosquito population
is very large or the host population is low (i.e. the vector-to-host ratio is high), as can occur with pronounced
seasonality, then the number of bites on hosts can be limited by the density of hosts. In this case, the number of bites
on a host per mosquito is close to $\sigma_h^k \hat N^k_h/N^k_v$ and the number of bites per host is close to
$\sigma_h^k$ \cite{chitnis2007bifurcation,chitnis2008determining}.

Similarly, $b^k_h = b_h(N^k_v,\hat N^k_h)=b^k/\hat N^k_h$ is the average number of bites per host per unit time and,
although we assume that, on average, all the mosquitoes have the same biting rate, not all of the hosts are being bitten
at the same rate.  The average number of bites per host per unit time in patch $k$ engaged in activity $j$ per unit time
is $b^k_{h,j} = \alpha_j b^k_h$, so that $\sum_j b^k_{h,j}N_{h,j}^k = \sum_j \alpha_jb_h^kN_{h,j}^k= b_h^k\hat N_h^k =
b^k$ and the total number of bites in the patch are preserved.

\subsubsection{Infection Rates}

Susceptible mosquitoes are infected at a rate $\lambda^k_v(t)$ defined as the product of the number of bites one
mosquito has per unit time, $b^k_v$; the probability of disease transmission (per bite) from an infectious host to the
mosquito, $\beta_{vh}$; and the average probability that the bitten host is infectious.  The probability, $\beta_{vh}$,
of contracting the pathogen after biting an infectious host depends upon the infectiousness of the host and the
susceptibility of the mosquito.  If the bites in the patch are uniformly distributed among all the hosts in the patch,
then the probability that the bitten host is infectious is $I_h/N_h$.  However, if the probability of being bitten
depends upon the activity the individual is engaged in, then the biting rate is not homogeneously distributed and the
average probability that the bitten host is infectious is proportional to $\hat I^k_h/\hat N^k_h$, where the scaled
infected population in the patch is defined as $\hat I^k_h=\sum_j \alpha_j I_{h,j}^k$.  Multiplying these three factors
together, we have the infection rate for mosquitoes is
\begin{equation}\label{E:lambdav}
\lambda^k_v = b^k_v \cdot \beta_{vh} \cdot \left( \frac{\hat I^k_h}{\hat N^k_h} \right)~~.
\end{equation}
This rate is used directly in model (\ref{E:mosquito2}), thereby coupling the ABM and the mosquito ODE model.

The rate at which hosts are infected from infectious mosquitoes is the product of the number of bites a typical host
engaged in activity $j$ gets per unit time, the probability that a bite is from an infectious mosquito, and the
probability of successful infection given a bite from an infectious mosquito. This rate can be expressed as
\begin{equation} \label{E:lambdah}
\lambda^k_{h,j}(t)= (\alpha_j b_h^k)\beta_{hv}\left(\frac{I^k_v}{N^k_v} \right)=b^k_{h,j} \cdot \beta_{hv}  \cdot  \left(\frac{I^k_v}{N^k_v} \right)~~.
\end{equation}
This is approximately the risk, per unit-time, that a susceptible person engaged in activity $j$ in patch $k$ will be
infected.
 
ABMs typically use discrete time steps $\Delta t$, which are modeled using Markov Chain techniques.  The infection rate
$\lambda^k_{h,j}(t)$ must be converted into a probability, $p^k_j$, that a susceptible person in patch $k$ engaged in
activity $j$ becomes infected in a time step.  We assume the time to infection is exponentially distributed, so the
probability of infection at the end of the time interval, $\Delta t$, given that the individual was uninfected at the
beginning of the time interval, is $1$ minus the probability that the person is not infected, or
\begin{equation} \label{E:probinf}
p^k_j= 1 - e^{-\lambda^k_{h,j} \Delta t }~~.
\end{equation} 
This probability is used by the ABM to determine whether a susceptible agent becomes infected in a given time step.  The
transition equations, and probability of transition, for the state of an agent are:
\begin{subequations}\label{E:agent}
  \begin{align}
    S_h &\mapsto E_h : Pr =  1 - e^{- \Delta t  \lambda^k_{h,j}} \\
    E_h &\mapsto I_h : Pr =  1 - e^{- \Delta t \nu_h}   \\
    I_h &\mapsto R_h : Pr =   1 - e^{- \Delta t \mu_h} .
  \end{align}
\end{subequations}
If there are $S^k_j$ susceptible hosts in activity $j$ in patch $k$, then on each time-step we generate a random number
$r \in [0, 1]$ for each susceptible host and declare the host infected if $r<p^k_j$.  This is a simple, albeit
computationally inefficient, approach for infecting agents. In future work, we will describe more computationally
efficient methods for updating an agent's infection status.
 
\subsection{Reproduction Number and Analysis}
\label{R0section}

For ABMs modeling directly transmitted disease, the basic and effective reproduction numbers can often be computed
exactly, since the disease status and who infected whom is often known for all agents. However, for the network patch
model, we do not keep track of individual mosquitoes, so we cannot say exactly how many mosquitoes are infected by a
single host, and of those mosquitoes exactly how many hosts are infected as a result. We can, however, start with a
simple heuristic based on averages for calculating the effective reproduction number in each patch.

For the fully homogenized model, the basic reproduction number can be computed exactly via two quantities
\cite{chikdengue,chitnis2013modelling}. The first is the average number of hosts infected by one infectious mosquito
introduced into a fully susceptible population, $R_{hv}$, and the second is the average number of mosquitoes a single
infectious host would infect if introduced into a full susceptible population, $R_{vh}$. These quantities can then be
multiplied together to form the type reproduction number, $\mathcal{R}_0^T=R_{vh}\cdot R_{hv}$, which is the average
number of hosts that would be infected (via mosquitoes) from one infected host introduced into a fully susceptible
population. The basic reproduction number, or number of new cases in the next generation from one infected individual
introduced in a fully susceptible population is $\mathcal{R}_0 = \sqrt{R_{vh}R_{hv}}$. For mosquito-borne disease, the
next generation for an infected host is infected mosquitoes and vice versa. The type reproduction number can be written
for this model formulation as $\mathcal{R}_0^T = (\mathcal{R}_0)^2$.

The \textit{effective} reproduction number measures the average number of secondary cases resulting from an infected
individual introduced into the population at any time point during an epidemic. It accounts for reduced susceptibility
of a population over time as individuals become immune or are vaccinated. When a disease has reached its endemic
equilibrium, the effective reproduction number is equal to one. The effective reproduction number can be approximated by
multiplying the basic reproduction number by the proportion of the population that is currently susceptible (i.e. not
infected or immune). We can then get a rough estimate for the effective reproduction number from host to vector,
$R_{vh}^k(t)$, and the effective reproduction number from vector to host, $R_{hv}^k(t)$, for the network patch model at
time $t$ in patch $k$ as
\begin{align}
R_{hv}^k(t) &= \left(\frac{\nu_v}{\mu_v+\nu_v} \cdot \frac{\sigma_v}{\mu_v} \cdot \frac{\sigma_h \hat N_h^k}{\sigma_h \hat N_h^k + \sigma_v N_v^k} \cdot \beta_{hv}\right) \cdot \left(\frac{(\hat N_h^k - \hat I_h^k)}{\hat N_h^k}\right)\\
R_{vh}^k(t) &= \left(\frac{\nu_h}{\mu_h+\nu_h} \cdot \frac{\sigma_h}{\mu_h+\gamma_h} \cdot \frac{\sigma_v N_v^k}{\sigma_h \hat N_h^k + \sigma_v N_v^k} \cdot \beta_{vh}\right) \cdot \left(\frac{S_v^k}{N_v^k}\right)
\end{align}
where the patch $k$ superscript for the variables and parameters is suppressed. The terms $\hat N_h^k$, $\hat I_h^k$,
$N_v^k$, and $S_v^k$ vary with time. However, the mosquito and disease parameters may vary with time as well. A full
explanation for each of the non-dimensional terms in $R_{hv}^k(t)$ and $R_{vh}^k(t)$ can be found in \cite{chikdengue}.
Over short periods of time, the product $R_{hv}^k(t)\cdot R_{vh}^k(t)\approx \mathcal{R}^{T,k}(t)$ will give us an
estimate for the effective type reproduction number at time $t$ in patch $k$ for the network patch model.

Methods for analyzing and quantifying the results of ABMs for disease spread are still a relatively new area and there
is much to be done in standardizing analysis of hybrid models combining ABMs with differential equations. The effective
reproduction number heuristic shown here gives an estimate for the potential of disease spread given pathogen
introduction, while the risk curve, $p^k(t)$, shows actual risk of acquiring disease in patch $k$ at that specific time
in a particular run.

\section{Network-patch Example}

We adapt an ABM for host movement and directly transmitted disease to a mosquito-borne disease in order to compare the
effects of host movement and environmental heterogeneity on mosquito-borne disease spread.  We describe the model
coupling method and evaluate the model in the context of disease modeling by comparing output for a well-mixed baseline
scenario to the corresponding standard ODE model for both humans and mosquitoes. We then illustrate how the
network-patch model differs from the associated homogeneous mixing model, even when all the hosts have the same activity
or exposure parameter.

Each location/node is assigned to a mosquito patch, within which the mosquitoes are assumed to be homogeneously
distributed.  We define the patch node density as the fraction of total nodes that are in a particular patch.  For this
example, we assume that all of the nodes have the same relative risk of mosquito exposure, meaning there is only one
`activity', $j=1$, the associated exposure risk is $\alpha_j=1$, and this risk depends only upon the mosquito population
in the patch associated with each node.  Therefore, the risk of a person being infected, or infecting mosquitoes,
depends upon the average force of infection, $\lambda_v^k$ and $\lambda_h^k$, for the $k^{th}$ patch.

The network agent-based and mosquito differential equation patch models must share information to calculate the force of
infection as functions of the biting rate, the fraction of mosquitoes infected, and the scaled fraction of hosts
infected in each patch.  This minimal communication between the network and patch models allows the flexibility to
easily adapt the approach to complex agent based models, such as EpiSimS \cite{mniszewski2008pandemic}. It is important
to make sure that the time step used in the ABM is also the time step used to progress the mosquito patch model.  Thus,
we feed the total number of agents, total number of infected agents, their associated exposure risks, and the time step
being used to the mosquito patch model each time it is updated. The mosquito patch model then returns a baseline risk of
being bitten by an infectious mosquito that can be scaled by each agent or location's exposure risk. See Figure
\ref{F:flowchart} for a flowchart of the ABM-patch model coupling method.

For model output, we tracked the proportion of agents infected at each patch for every time step. We also tracked the
patch each agent became infected in so we could measure the contribution to overall infection burden of each
patch. Finally, we tracked the risk, or $p_j^k=p^k$, in each patch as well as the effective and basic reproduction
number. Note that the risk, $p_j^k$, depends on the time step chosen, mosquito habitat patch, relative density of
mosquitoes and humans, and the proportion of mosquitoes in the patch that are currently infected.

\subsection{Simulations}

We consider a network as illustrated in Figure \ref{F:NetworkPatch} for disease progression and mosquito parameters
appropriate for dengue.  We assume that there are three patches with distinct mosquito habitat and dynamics determined
by landscape, weather, available hosts, and breeding sites. We neglect mosquito movement between patches which is a
reasonable assumption for \textit{Aedes aegypti} mosquitoes, common vectors of dengue.  We first establish a baseline
case in order to compare different modeling assumptions as we incorporate heterogeneity. The model parameters (Table
\ref{T:param}) for this baseline case are the same in each patch, and are constant in time.  In the baseline case, each
patch has the same density of mosquitoes and humans with the same resulting vector-to-host ratio.  A high human movement
rate was used for the baseline case to approximate a well-mixed human population.  We simulated the baseline scenario
100 times to approximate the distribution of possible solutions created by the inherent stochasticity of the ABM. We
solved the approximate mean-field ODE equations for the network-patch model using the mean values for human disease
progression parameters from the ABM.  The resulting ODE model for both humans and mosquitoes is the Manore et al. model
\cite{chikdengue} for dengue and chikungunya.

Figure \ref{F:Baseline2} compares an ensemble of simulations for the baseline network-patch model with the ODE
mean-field model.  Although the distribution of solutions to the stochastic ABM are the same in each patch, we observe
some differences in the samples.  The mean-field ODE model for humans and mosquitoes is a good approximation of the
stochastic network-patch model in this highly-mobile, well mixed population with identical mosquito habitat patches.  As
we expected, our network-patch model matches the standard differential equation models well with some variation around
the mean-field approximation. This allows for meaningful comparison of the hybrid model with the standard non-spatial
ODE models that do not account for individual movement.

\subsection{The Impact of Heterogeneity}
\label{het}

Next, we included heterogeneity in patch parameters and in vector-to-host ratios as described in Figure
\ref{F:NetworkPatch}. Patch 1 (green) is assumed to have locations including mostly buildings with AC and screens and
not to have many mosquito breeding sites. Mosquitoes here have relatively low egg-laying and survival rates on average,
so there is lower risk in this patch. Patch 2 (blue) has more human-made and/or natural breeding sites and fewer
buildings with AC and screens. Thus, patch 2 has more humans at risk for mosquito bites. Mosquito populations in patch 2
are more robust than in patch 1 with medium density and an associated medium risk level. Patch 3 (red) is assumed to be
prime mosquito habitat, having, on average, high mosquito density. Mosquito contact with humans in patch 3 is high, with
locations consisting mostly of open air dwellings, buildings without screens, and outdoor locations.  We assumed that
patch 1 has the highest number of humans, patch 2 medium number of humans, and patch 3 the fewest humans.

\begin{table}
  \caption{{\bf Patch Parameters:} The parameter values for the simulations experiments in Figures
    \ref{F:totalinfect}-\ref{F:peaknumber}. Notice, for the simulations in the heterogeneous scenarios only the movement
    rate changes.}
  {\begin{tabular}{|l|l|l|}
      \hline
      Parameter & Value (P1, P2, P3) & Explanation \\
      \hline
      \multicolumn{3}{|l|}{\textit{Baseline} } \\
      \hline
      $\sigma_h$ & (19, 19, 19) & Maximum bites on a human per day \\
      $K_v$ & (1000, 1000, 1000) & Mosquito carrying capacity \\
      Patch Density & $\left(\frac{1}{3}, \frac{1}{3}, \frac{1}{3}\right)$ & Fraction of locations per patch \\
      Movement Rate & $\ln \mathcal{N}(\mu, \sigma^2)$, mean $= 1$, var. $= 0.001$ &  Average number of location changes per day \\
      \hline
      \multicolumn{3}{|l|}{\textit{Heterogeneous Patch, High movement} } \\
      \hline
      $\sigma_h$ & (5, 19, 30) & Maximum bites on a human per day \\
      $K_v$ & (750, 1500, 3750) & Mosquito carrying capacity \\
      Patch Density & $\left(\frac{1}{2}, \frac{1}{3}, \frac{1}{6}\right)$ & Fraction of locations per patch \\
      Movement Rate & $\ln \mathcal{N}(\mu, \sigma^2)$, mean $= 1$, var. $= 0.001$ & Average number of location changes per day \\
      \hline
      \multicolumn{3}{|l|}{\textit{Heterogeneous Patch, Medium Movement}} \\
      \hline
      $\sigma_h$ & (5, 19, 30) & Maximum bites on a human per day \\
      $K_v$ & (750, 1500, 3750) & Mosquito carrying capacity \\
      Patch Density & $\left(\frac{1}{2}, \frac{1}{3}, \frac{1}{6}\right)$ & Fraction of locations per patch \\
      Movement Rate & $\ln \mathcal{N}(\mu, \sigma^2)$, mean $= 0.1$, var. $= 0.001$ & Average number of location changes per day \\
      \hline
      \multicolumn{3}{|l|}{\textit{Heterogeneous Patch, Low Movement} } \\
      \hline
      $\sigma_h$ & (5, 19, 30) & Maximum bites on a human per day \\
      $K_v$ & (750, 1500, 3750) & Mosquito carrying capacity \\
      Patch Density & $\left(\frac{1}{2}, \frac{1}{3}, \frac{1}{6}\right)$ & Fraction of locations per patch \\
      Movement Rate & $\ln \mathcal{N}(\mu, \sigma^2)$, mean $= 0.01$, var. $= 0.001$ & Average number of location changes per day \\
      \hline
      \multicolumn{3}{|l|}{\textit{Heterogeneous Patch, Very Low Movement} } \\
      \hline
      $\sigma_h$ & (5, 19, 30) & Maximum bites on a human per day \\
      $K_v$ & (750, 1500, 3750) & Mosquito carrying capacity \\
      Patch Density & $\left(\frac{1}{2}, \frac{1}{3}, \frac{1}{6}\right)$ & Fraction of locations per patch \\
      Movement Rate & $\ln \mathcal{N}(\mu, \sigma^2)$, mean $= 0.001$, var. $= 0.001$ & Average number of location changes per day \\
      \hline
\end{tabular}}
\label{T:PatchParams}
\end{table}

\begin{table}
\caption{{\bf Patch Parameters:} The parameter values used for all numerical experiments.}
{\begin{tabular}{|l|l|l|}
\hline
\multicolumn{3}{|l|}{\textit{All Experiments} } \\
\hline
Parameter & Value (P1, P2, P3) & Explanation \\
\hline
 $\psi_v$ & (0.3, 0.3, 0.3) & Emergence rate of female mosquitoes \\
 $\sigma_v$ & (0.5, 0.5, 0.5) & Max mosquito bite demand per day \\
 $\beta_{hv}$ & (0.33, 0.33, 0.33) & Probability of M-to-H transmission \\
 $\beta_{vh}$ & (0.33, 0.33, 0.33) & Probability of H-to-M transmission \\
 $\nu_v$ & (0.1, 0.1, 0.1) & Mosquito E-to-I rate \\
 $\mu_v$ & $\left(\frac{1}{14}, \frac{1}{14}, \frac{1}{14} \right)$ & Mosquito death rate \\
 $r_v$ & $\psi_v - \mu_v$ all patches & Intrinsic growth rate \\
\hline
\multicolumn{3}{|l|}{\textit{Patch Independent Parameters}} \\
\hline
 Total Number of Locations & 300 & Distributed among patches by density \\
 Edge Probability & 0.03 & Prob. two locations connect \\
 Total Human Pop. & 1500 & Distributed equally among locations \\
 Initial Infected \% & 0.5\% & \% initially infected per patch \\
 Recovery Rate & $\ln \mathcal{N}(\mu, \sigma^2)$, mean $= 1/6$, var. $= 0.001$ & Avg. human recovery of 6
 days \\
 Incubation Rate & $\ln \mathcal{N}(\mu, \sigma^2)$, mean $= 1/5$, var. $= 0.001$ & Avg. human E-to-I of 5
 days \\
\hline
 Total simulation time & 200 days & \\
 ABM time step & 0.25 days & \\
 Mosquito r-k time step & 0.005 days & \\
\hline
\end{tabular}}
\label{T:ConstantPatchParams}
\end{table}

The baseline version of the model for high human movement closely resembles results from a fully homogenized model with
ODEs for both humans and mosquitoes (Figure \ref{F:Baseline2}).  In this situation, justification for using the
network-patch model rather than an ODE or stochastic differential equation model may be limited. However, this does
provide a good baseline for comparison with other models and with the following scenarios that incorporate heterogeneity
in patch parameters and host density and movement.

We considered three different host movement scenarios for heterogeneous habitat patches. Scenario 1 was high host
movement between nodes, scenario 2 medium movement, scenario 3 low movement, and scenario 4 very low host movement
(results not shown).  Mosquito parameter values, human density, and mosquito density varied between the three
patches. The highest human density and lowest mosquito density is in Patch 1 (green), medium human density and medium
mosquito density in Patch 2 (blue) and lowest human density and highest mosquito density in Patch 3 (red). Thus, Patch 3
has the highest vector-to-host ratio and Patch 1 has the lowest vector-to-host ratio.  We ran each of the scenarios 100
times to capture inherent stochasticity in the ABM.  Stochasticity in placement of initial infectious humans and agent
movement between patches is much more visible for the heterogeneous scenarios.  Parameter values used for the runs can
be found in Tables \ref{T:PatchParams} and \ref{T:ConstantPatchParams}.

In the high host movement rate scenario, risk varies dramatically between patches. Since Patch 3 (red) had a high
relative density of mosquitoes, it also had the highest patch risk after the epidemic became established (Figure
\ref{F:High}, bottom row).  Furthermore, more hosts became infected in Patch 3 than in any other patch (Figure
\ref{F:High}, middle row).  However, because of the high movement rates, if the population in each patch was sampled at
a time point, one would not see higher prevalence in the highest risk patch. Therefore, it would be difficult to
determine that the high risk patch should be targeted for mitigation in the presence of limited public health or
mosquito control resources based on human prevalence data alone.

For the medium host movement rate scenario, agents are less likely to visit every patch often, so stochasticity in the
initial location of infected agents and their subsequent movement has a greater effect on the overall dynamics (Figure
\ref{F:Medium}). Also, once an agent is in a patch, it is more likely to stay there since the probability of moving is
lower.  Prevalence in the high risk patch is higher than in any other patch once the epidemic takes off.  In this
scenario, unlike the high movement scenario, the location of the high risk patch could generally be determined by
prevalence data.

In the low (Figure \ref{F:Low}) and very low (results not shown) host movement rate scenarios, we see much more
heterogeneity between epidemics in the habitat patches. For example, the epidemic rarely peaks at the same time in the
patches.  This can be good from a control point of view, because once an epidemic is noticed in one patch, measures can
be taken to reduce risk of spreading to the other patches. Prevalence is always higher in the high risk patch for the
low movement scenario. Still, in the low movement case, the epidemic usually moves to every patch over the 200 day
simulation time.  For very low movement rates, the patches behave like separate villages with only sporadic connection
between them. In the very low movement case, an ODE patch model with low stochastic movement rates between patches could
probably capture the dynamics as well as the network-patch hybrid model. The network patch model shows increasing
sensitivity to the initially infected individual and their subsequent movement as patch heterogeneity increases and host
movement rates are lower and more sporadic.

\begin{figure}[h]
  \begin{center}
   \includegraphics[scale=0.5]{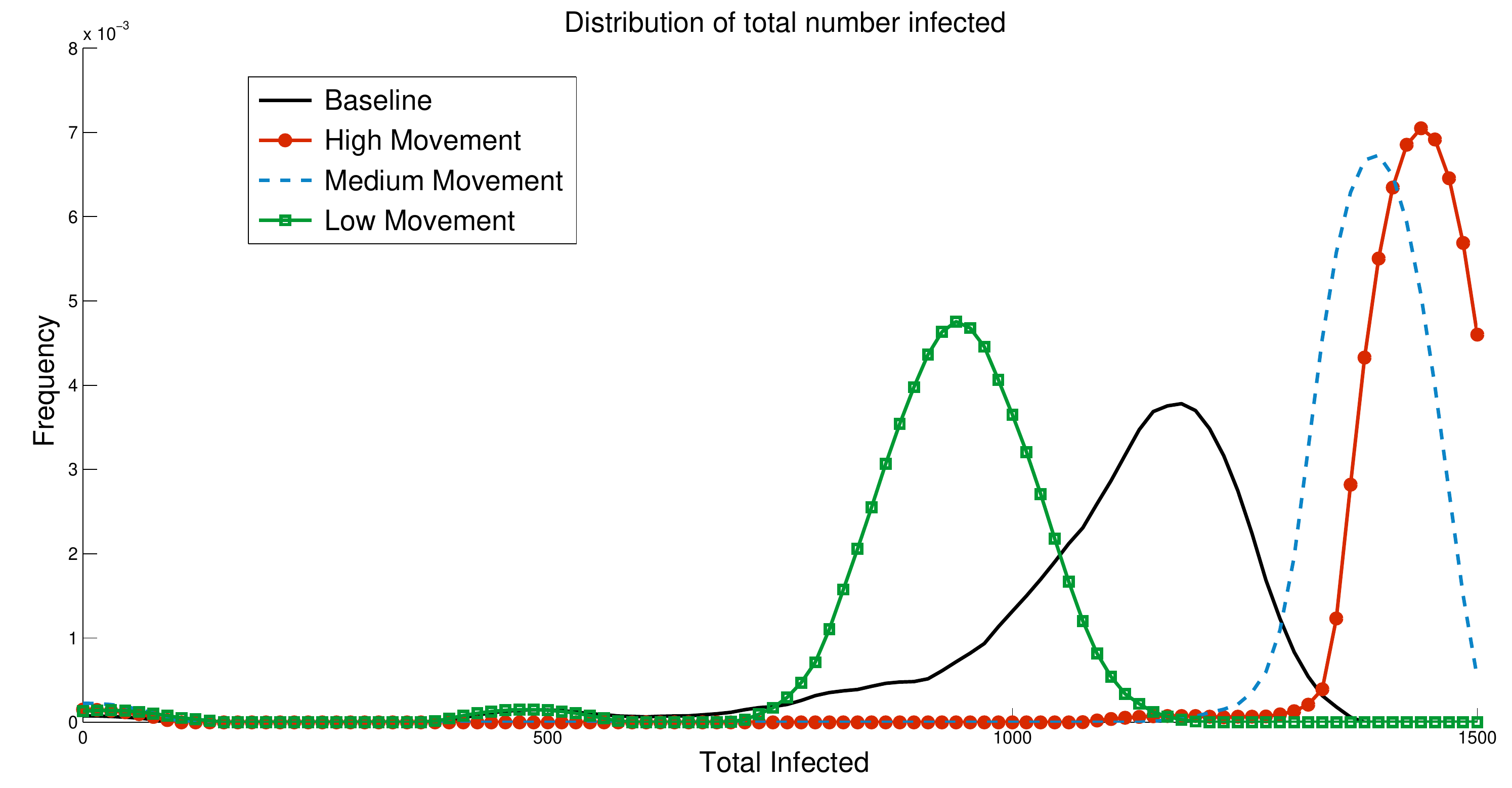}
  \end{center}
  \caption{ Distribution of the total number of people infected over the course of the simulation (200 days) for each
    scenario. Each scenario was run 100 times to capture intrinsic uncertainty due to stochasticity. The pathogen is
    introduced into a fully susceptible population of 1,500 hosts with no mitigations implemented. In the heterogeneous
    scenarios with one high risk patch and high or medium human movement, the total consequence is higher than for the
    baseline homogeneous scenario. However, with low movement between patches, the scenario with one high risk patch
    results in lower total consequence than the baseline.  }
  \label{F:totalinfect}
\end{figure}

\begin{figure}[h]
  \begin{center}
     \includegraphics[scale=0.52]{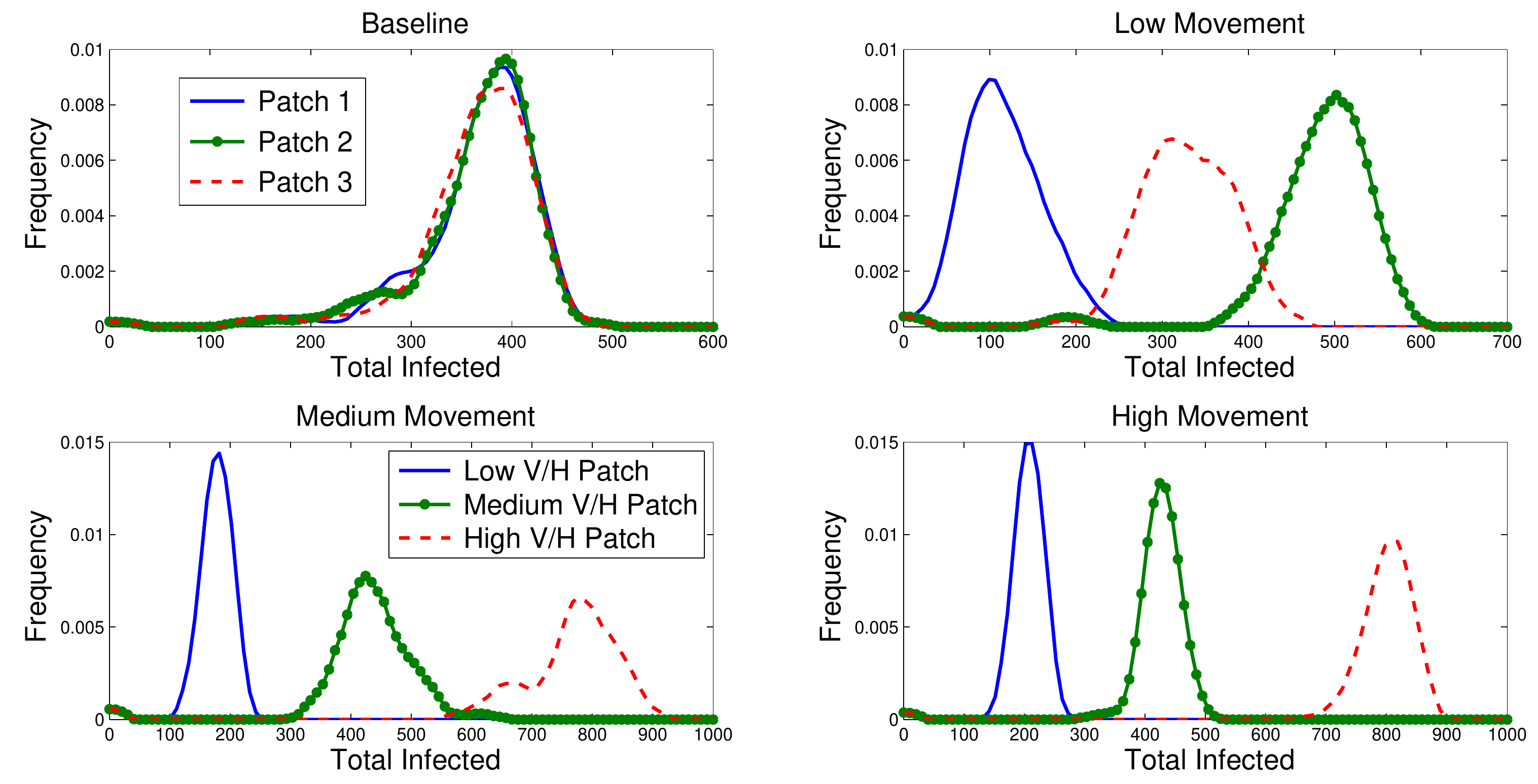}
  \end{center}
  \caption{ Distributions of the total number of hosts \textbf{initially infected} in each patch for the different
    scenarios. For the baseline case, all patches have the same density of mosquitoes and each patch is responsible for
    approximately the same number of initial infections. For the heterogeneous scenarios, red dashed is the high risk
    patch, green dotted the medium risk and blue solid the low risk patch. For high and medium host movement, the
    highest risk patch is responsible for the most infections. For the low movement scenario, the medium risk patch is
    responsible for the most infections and the low risk patch is responsible for the fewest. This difference from the
    high/medium movement scenarios can be explained by the fewer number of resident hosts in the high risk patch. Since
    movement between patches is low in the low movement scenario, the high risk patch runs out of susceptible hosts
    faster.}
  \label{F:totalpatchinfect}
\end{figure}

\begin{figure}[h]
  \begin{center}
     \includegraphics[scale=0.5]{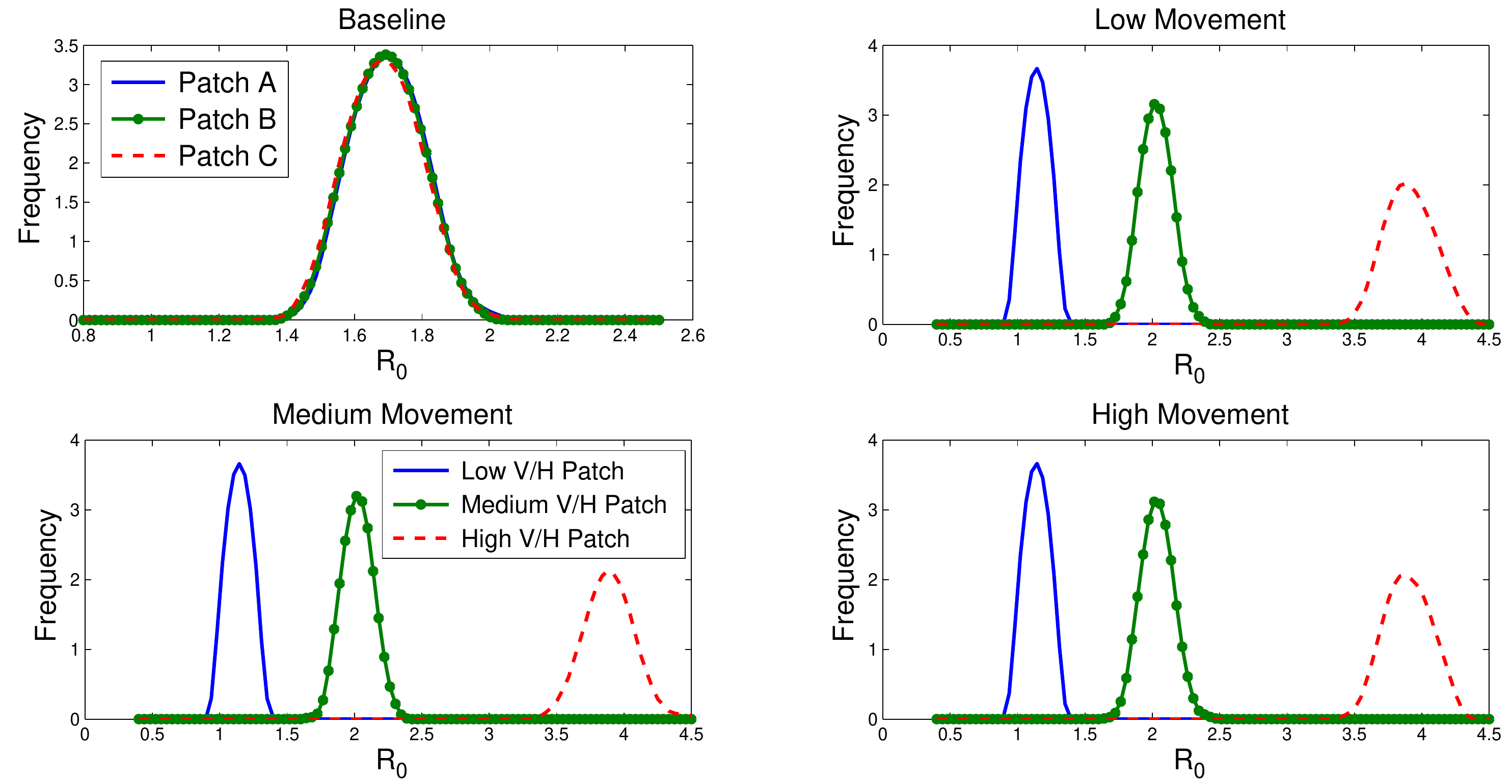}
  \end{center}
  \caption{ Distributions for the estimated basic reproduction number for each patch. For the baseline case, $R_0\approx
    1.7$ while in the heterogeneous cases, in the low risk patch $R_0$ is just above $1$, in the medium risk $R_0$ is
    approximately $2$ and in the high risk patch, $R_0$ is just under $4$. Notice that the basic reproduction number
    distribution (estimated as the effective reproduction number computed at the first time step for each run) is very
    similar among the heterogeneous scenarios. However, heterogeneity in movement patterns between patches results in
    different total consequence for each scenario as seen in Figure \ref{F:totalpatchinfect}. }
  \label{F:R0dist}
\end{figure}

\begin{figure}[h]
  \begin{center}
     \includegraphics[scale=0.5]{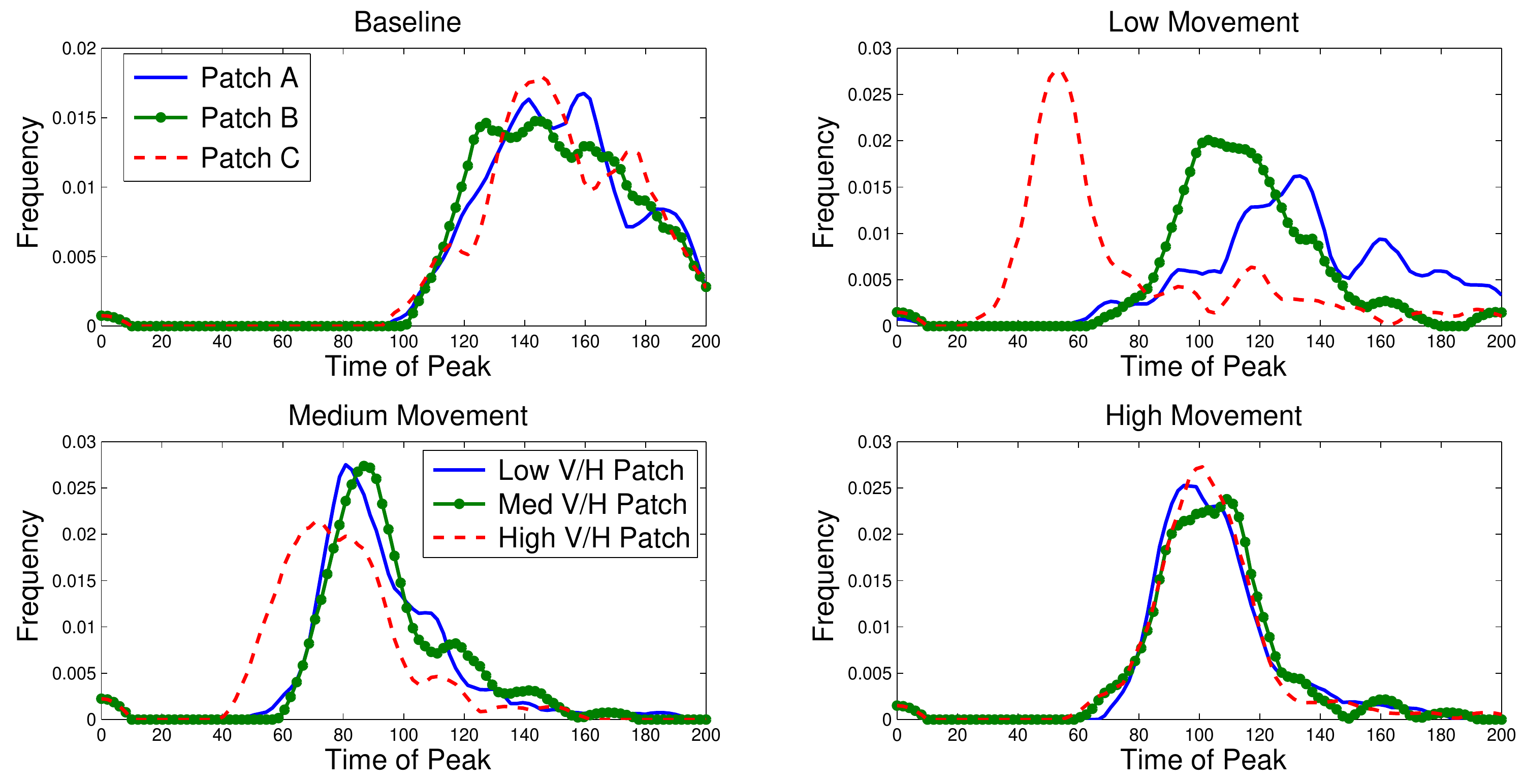}
  \end{center}
  \caption{ Distributions for the timing of the peak of the epidemic in each patch. For the low and medium movement
    scenarios, the high risk patch peaks before the other patches in general. For the high movement case and the
    baseline case, the patches reach the epidemic peak at approximately the same time. The low movement scenario has the
    most variation is epidemic timing. }
  \label{F:peaktime}
\end{figure}

\begin{figure}[h]
  \begin{center}
     \includegraphics[scale=0.5]{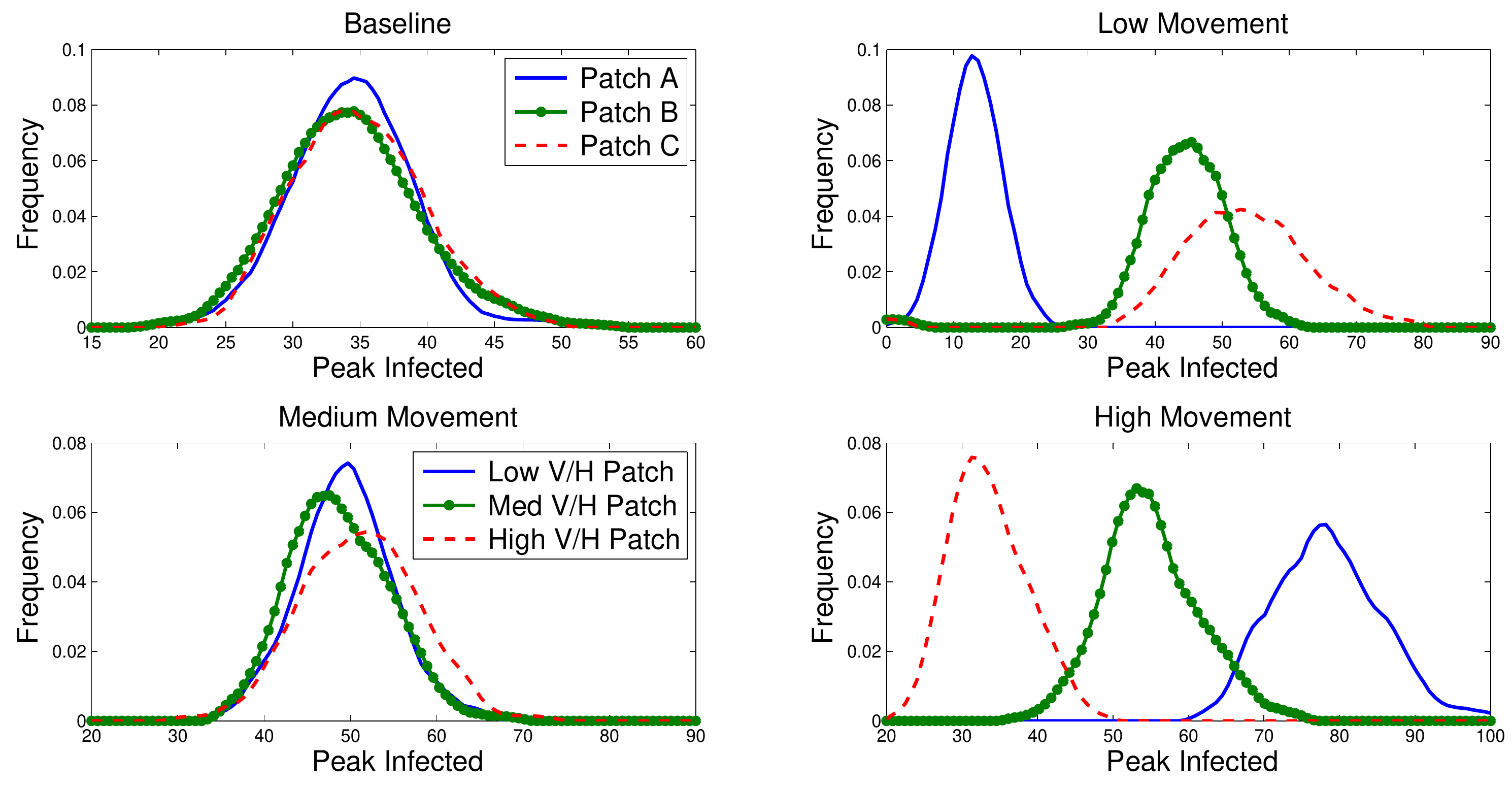}
  \end{center}
  \caption{ Distributions for the number of people who are infectious at the time of the epidemic peak in each
    patch. This is highly dependent on both patch risk and the movement patterns between the patches. The baseline and
    medium movement scenarios are the most similar for this metric, while the low and high movement rates have opposite
    patterns. This is again a reflection of the tradeoff between the risk of a patch from high vector density and the
    number of hosts in the patch and/or accessibility of the patch to hosts moving in and out.}
  \label{F:peaknumber}
\end{figure}

The overall consequence is higher for the heterogeneous, high and medium human movement scenarios than for the baseline
scenario (Figure \ref{F:totalinfect}) because of the presence of a high risk patch that is visited often by most agents
due to the higher movement rates. We see that for the baseline scenario, each patch is responsible for initially
infecting about the same number of hosts (Figure \ref{F:totalpatchinfect}). However, for the heterogeneous scenario, the
high risk patch is responsible for a relatively high number of initial infections, suggesting again that a targeted
response may be effective. This is despite the fact that the distribution of the estimated basic reproduction number in
each patch is the same for all movement scenarios (Figure \ref{F:R0dist}).  There is also more variation in timing of
the epidemic peak across the runs for the heterogeneous scenarios (Figure \ref{F:peaktime}). For the high movement
scenario, there is enough host movement that the peak of the epidemic is at approximately the same time in all
patches. However, for the low and medium movement scenarios, the epidemic tends to peak in the high risk patch
first. The peak number of infectious people residing in each patch at the peak of the epidemic is shown in Figure
\ref{F:peaknumber} again highlighting the tradeoff between the risk of a patch from high vector density and the number
of hosts in the patch and/or accessibility of the patch to hosts moving in and out. The figures illustrate that the
network-patch model will capture heterogeneity that would not be captured by a differential equation model for humans
and mosquitoes.

These simulation results show that heterogeneity in host movement and spatial heterogeneity in mosquito density can play
an important role in the spread, timing and size of mosquito-borne pathogen epidemics. The network-patch model
can capture this heterogeneity using the power of ABMs already tuned to particular host behavior and landscapes. The
importance of spatial and behavioral heterogeneity has also been observed for directly transmitted pathogens and in some
current studies on mosquito-borne viruses such as dengue \cite{barmak2011dengue,cosner2009effects}. Capturing this
inherent heterogeneity can be important for biosurveillance, mitigation, and treatment during outbreaks. This motivates
the need for adapting ABMs with detailed host activity, behavior, social, demographic, and geographical data to
mosquito-borne diseases as a step toward creating real time and spatially explicit risk maps and mitigation strategies.

\subsection{Mosquito heterogeneity across time and species}

\begin{figure}[h]
  \begin{center}
   \includegraphics[scale=0.5]{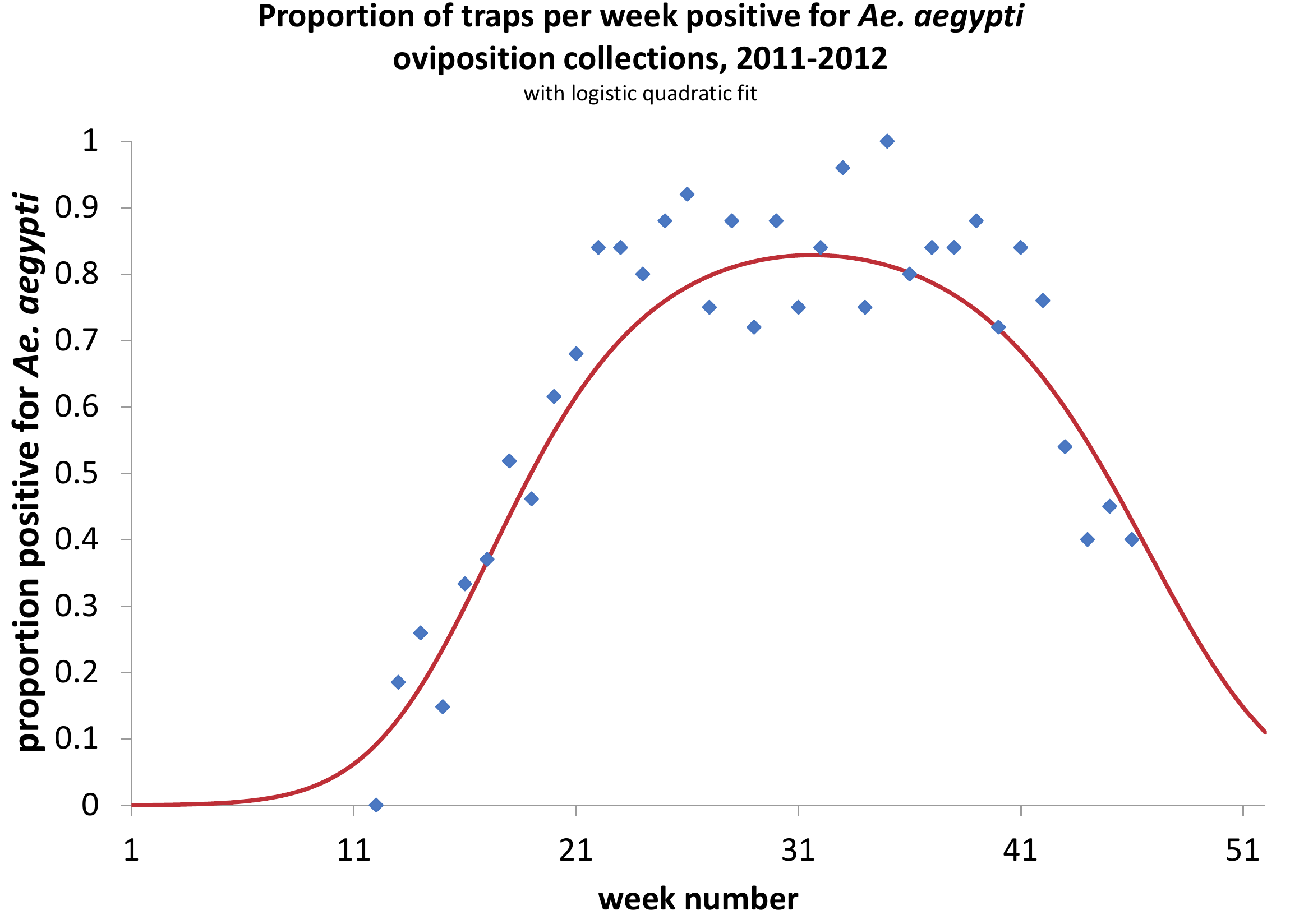}
  \end{center}
  \caption{Example of seasonality in mosquito populations from New Orleans mosquito trap data
    \cite{wessondavis2014}. Mosquito carrying capacities, emergence and/or death rates can be adjusted to follow
    seasonal patterns. }
  \label{F:season}
\end{figure}

The illustrative example given above is for dengue spread in humans by one mosquito species, \textit{A. aegypti}. We
assumed constant density of mosquitoes in each patch across time. However, data about seasonal variation mosquito
density such as in Figure \ref{F:season} can be incorporated into the mosquito patch model by varying mosquito carrying
capacities, $K_v^k(t)$, emergence rates, $\Phi_v^k(t)$, and/or death rates, $\mu_v^k(t)$ with time. Methods such as
parameter fitting or data assimilation can be used to determine how the parameters change with time, temperature,
policy, etc.  Although we used humans movement to illustrate the method, an ABM for cattle movement between farms and/or
wild animal movement between habitat could be adapted to consider Rift Valley fever infection. Bird movement in urban
habitat could couple with the mosquito patch model to model West Nile virus risk.

The network-patch framework can also incorporate multiple mosquito species in the model. In a mixed mosquito species
scenario, one patch can have \textit{A. aegypti} mosquitoes while another has \textit{A. albopictus} or some
patches could contain both species. Including two mosquito species will be necessary to model emergence of chikungunya, which is more
efficiently transmitted by \textit{A. albopictus} than dengue is, thus necessitating inclusion of both \textit{A. albopictus} and \textit{A. aegypti}. The two species of mosquitoes differ in traits including
human biting rates and vector competence which can be captured by adapting the parameter values for the mosquito model
to each species. The network-patch model can incorporate overlap between the species by adding equations to the patch
model for an additional mosquito species so that areas in the city can contain both species simultaneously. For Rift
Valley fever or West Nile virus, both $Aedes$ and $Culex$ mosquitoes may be important and have very different dynamics
and responses to climate and land use. This illustrates the ability of the model to incorporate multiple layers of
mosquito habitat and populations to include different mosquito species and behavior.

\section{Discussion}

We describe a method for coupling an ABM for directly transmitted diseases in hosts with a mosquito patch model in order
to model mosquito-borne diseases while capturing important spatial, temporal, and behavioral heterogeneity. This
approach makes use of the considerable infrastructure already available in large ABMs and expands the scenarios under
which these models can be used. We used the model to explore mosquito-borne virus spread in heterogeneous environments
illustrating the utility of the network-patch approach.

The examples considered here show that heterogeneity in mosquito habitat and host movement can change the dynamics of
the initial spread and spatial patterns of a mosquito-borne disease. We investigated the importance of heterogeneity in mosquito population dynamics and host movement on pathogen
transmission, motivating the utility of detailed models of individual behavior and observed that the random mixing model only captured the dynamics of the the high movement rate scenario. Our hybrid agent-based/differential equation model is able to quantify the importance of the 
heterogeneity in predicting the spread and invasion of mosquito-borne pathogens. We observed that  the total number of infected people is greater in heterogeneous patch models 
 with one high risk patch and high or medium human movement than it would be in a random mixing  homogeneous model. 
However, when there is low movement between patches, the scenario with one high risk
patch resulted in lower total consequence than the baseline.   
Mitigation strategies can be more effective when guided by realistic models such as 
models outlined here that extend the capabilities of existing agent-based models to include vector-borne
diseases. Mitigation and prevention strategies can be
optimized with better understanding of interactions between space, climate, and host movement resulting in observed
heterogeneities. Future work will adapt ABMs such as the Epidemic Simulation System (EpiSimS)
\cite{mniszewski2008episims,eubank2004modelling,stroud2007} and Framework for Reconstructing Epidemiological Dynamics
(FRED) \cite{FRED} to model mosquito-borne diseases such as dengue and chikungunya in areas that are at risk.

The patch approach for creating dynamic risk maps by coupling host social network and movement models with the
environment can be adapted for other scenarios such as environmental contaminants or other spatially and temporally
varying hazards. Feasibility of using this approach for the specific phenomenon being modeled is determined by the
temporal and spatial scales at which the host agents move and progress through the disease as well as the
spatio-temporal scales that the environmental hazards change over, whether the environmental hazard is mosquitoes,
livestock, birds, other wildlife, etc.

In applying this framework one should keep in mind that the ODE model of risk represents a homogenization of a large
number of random events. This is applicable in mosquito transmitted diseases where the vectors can be approximated as
risk `clouds' at some scale, but may be less applicable in other situations. Additionally, it may be important to
incorporate stochasticity in the ODE model. In fact, the underlying patch model for the hazard being modeled could be of
many different forms, including discrete, statistical, or Markov chain models, provided it can communicate an
appropriate risk to the ABM. The mosquito population dynamics and disease model we chose as the basis for the patch part
of the model is general and can be adapted to multiple pathogens. However, the network-patch approach for coupling a
mosquito model to an ABM described in this paper would work well for most standard homogeneous mosquito-borne disease
models.

Examples of applications for the hybrid modeling framework include mosquito-borne pathogens such as dengue and
chikungunya, as well as pathogens spread by other arthropod vectors, such as lyme disease, with an appropriate
underlying model for ticks.  This method could also be adapted to zoonotic mosquito-borne diseases such as Rift Valley
fever or West Nile virus and could be coupled with a spatial model for livestock and/or wildlife. If movement and
heterogeneity among individual animals is important, the additional animal hosts could be incorporated as agents that
would couple to the mosquito model. If that level of granularity is not needed for the additional animal hosts, they
could be incorporated into the network-patch model as another layer of patches with ODEs (or similar well-mixing models)
governing population dynamics and disease transmission. For example, a model for West Nile virus could be implemented by
layering patches of bird habitat over a city and modeling bird dynamics and transmission by systems of differential
equations that are coupled with the mosquito patches. Movement of birds (or mosquitoes) between patches could be added
if needed.  Parameters of a patch could change with time as weather or other factors affecting patch dynamics change,
allowing for exploration of scenarios such as climate change. Patch-specific mitigation strategies as well as social
strategies when agents avoid a patch with high hazard rates can also be implemented.

In conclusion, coupling ABMs for hosts with the environment helps us to use existing tools to explore the role that
spatial heterogeneity and host movement play in the emergence and spread of infectious disease, particularly
mosquito-borne pathogens. We have presented a prototype for creating a dynamic risk map that changes with time as
mosquito dynamics and host behavior and movement change. In future work we will use methods arising in sensitivity
analysis and uncertainty quantification to determine the most important factors in disease spread as indicated by these
models. This will provide us with valuable insight into methods for disease control as well as lend important validation
for simulation techniques.

\section*{Acknowledgments}
CM and JH are partially supported through grants from the NIH/NIGMS grant in the Models of Infectious Disease Agent
Study (MIDAS) program, U01-GM097661-01 and CM is partially supported by the NSF MPS Division of Mathematical Sciences
NSF/MPS/DMS grant DMS-1122666 and by NSF SEES Fellow grant CHE-1314029. KH is partially supported by NIH/NIGMS MIDAS
grant U01-GM097658. We would like to thank Ben McMahon, Jeanne Fair, Rebecca Christofferson, Susan Mniszewski, Helen
Wearing, Geoffrey Fairchild, and other NIH-MIDAS colleagues for useful conversations and suggestions.

\bibliographystyle{plainnat}
\bibliography{NPM_tJBD}
\newpage

\newpage

\begin{figure}[h]
  \begin{center}
    \includegraphics[scale=0.7]{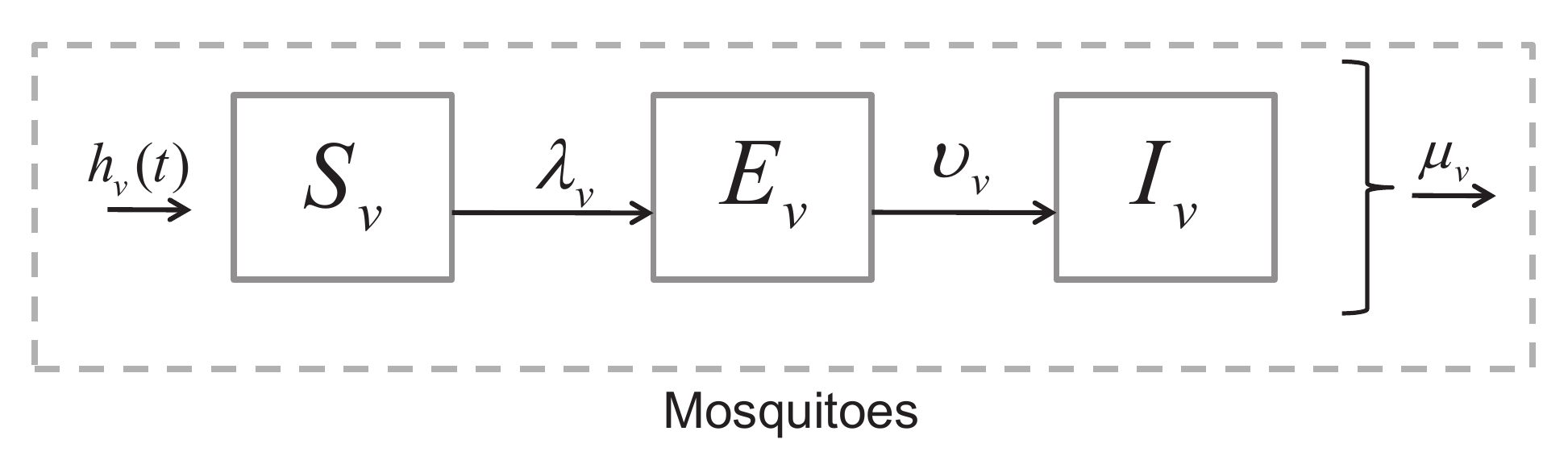}
  \end{center}
  \caption{In the female mosquito model, susceptible adults are infected at a rate $\lambda_v$ and pass through the
    exposed compartment, $E_v$, to the infectious compartment, $I_v$. All compartments contribute to reproduction, and
    we assume the death rate is independent of the infection status.}
  \label{F:mosqu}
\end{figure}

\newpage

\newpage

\newpage

\newpage

\newpage

\newpage

\newpage

\appendix

\setcounter{figure}{0}
\renewcommand{\thefigure}{A\arabic{figure}}
\section{Supporting material}

We present here three representative runs for each of the network-patch scenarios as well as a flow chart illustrating
the modeling coupling process. 

\newpage

\begin{figure}[h!]
  \begin{center}
     \includegraphics[scale=0.7]{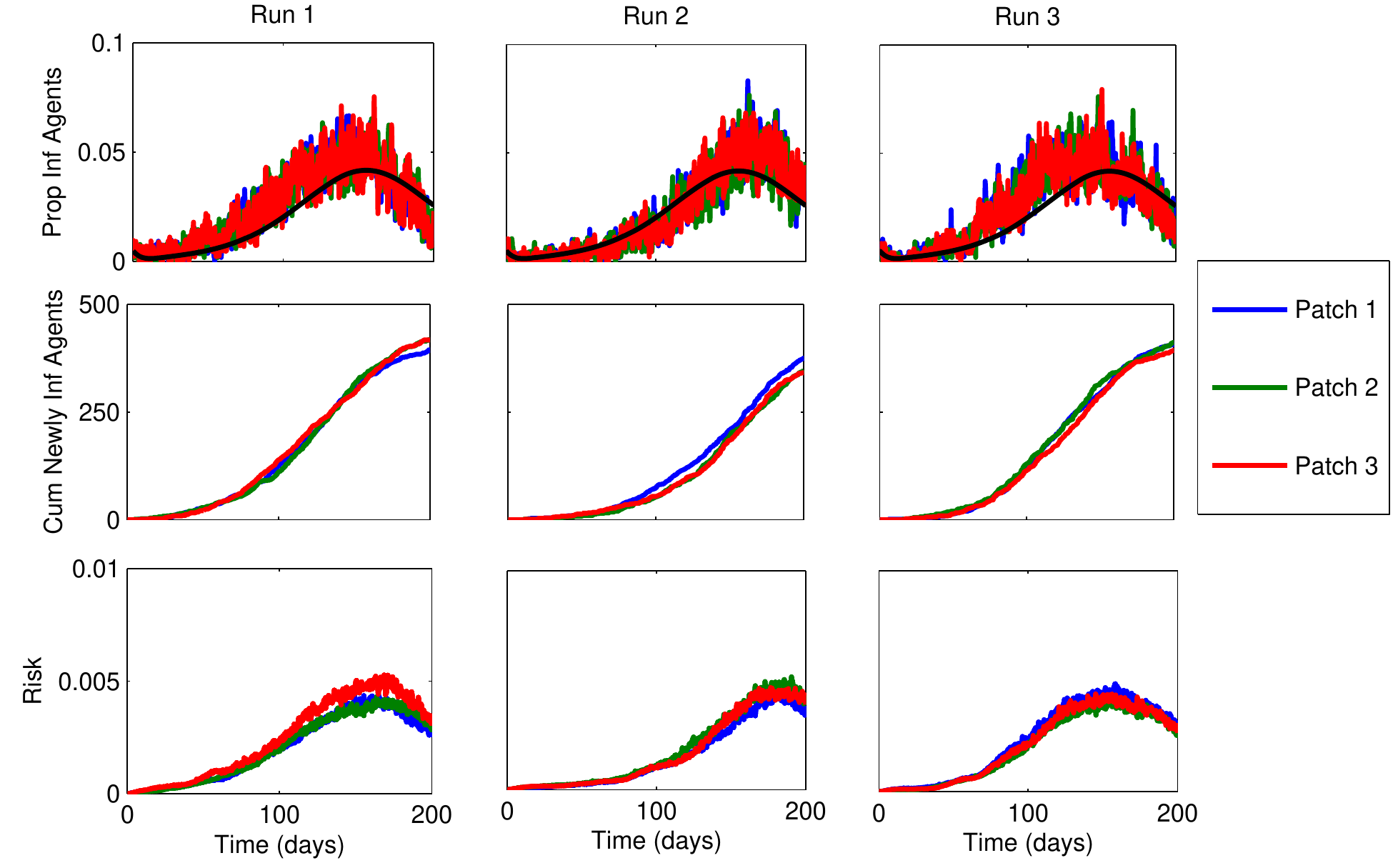}
  \end{center}
  \caption{ Three representative runs for the baseline scenario with the cumulative number initially infected in each
    patch and the risk in each patch over time. The risk can be thought of as the probability of an agent who was in the
    patch for 6 hours (for our simulations $\Delta t=0.25$ days) becoming infected during that time. The ODE model
    solution is represented by a solid black line. The baseline case matches well with the well-mixed ODE model.}
  \label{F:Baseline2}
\end{figure}

\newpage

\begin{figure}[h]
  \begin{center}
     \includegraphics[scale=0.7]{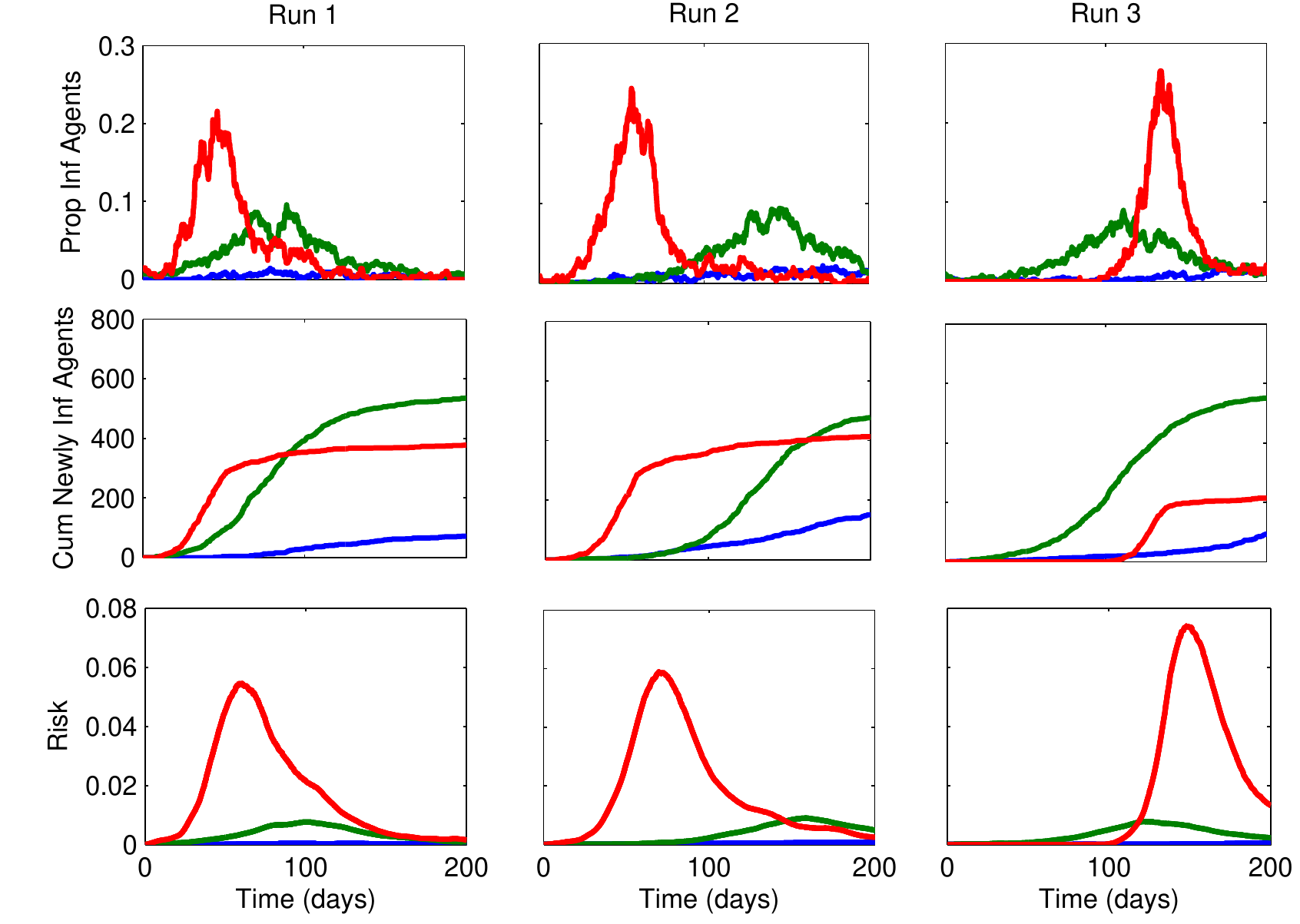}
  \end{center}
  \caption{ Three representative runs for the heterogeneous, low movement scenario with the cumulative number initially
    infected in each patch and the risk in each patch over time.}
  \label{F:Low}
\end{figure}

\newpage

\begin{figure}[h]
  \begin{center}
     \includegraphics[scale=0.7]{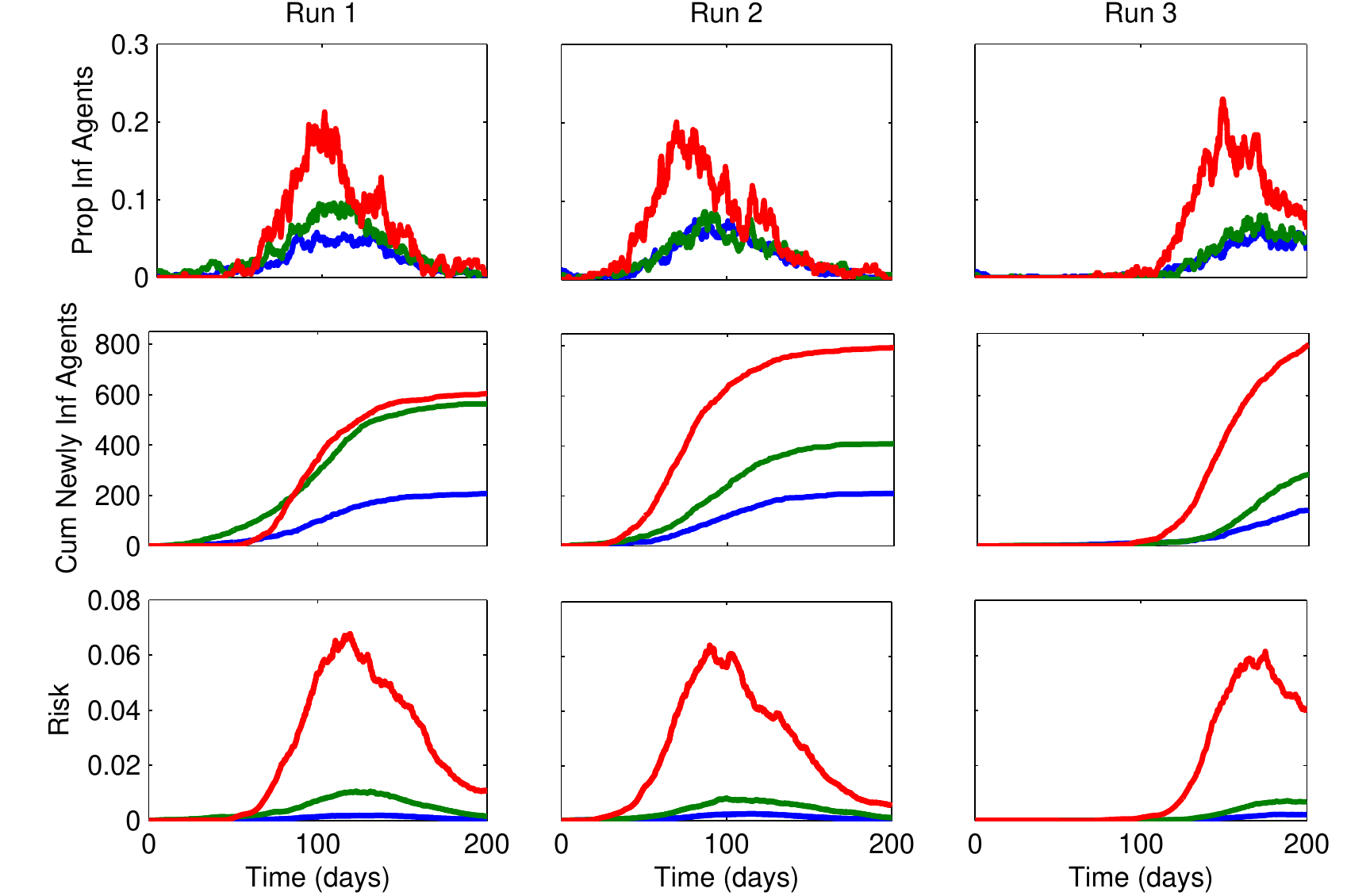}
  \end{center}
  \caption{ Three representative runs for the heterogeneous, medium movement scenario with the cumulative number
    initially infected in each patch and the risk in each patch over time.}
  \label{F:Medium}
\end{figure}

\newpage

\begin{figure}[h]
  \begin{center}
     \includegraphics[scale=0.7]{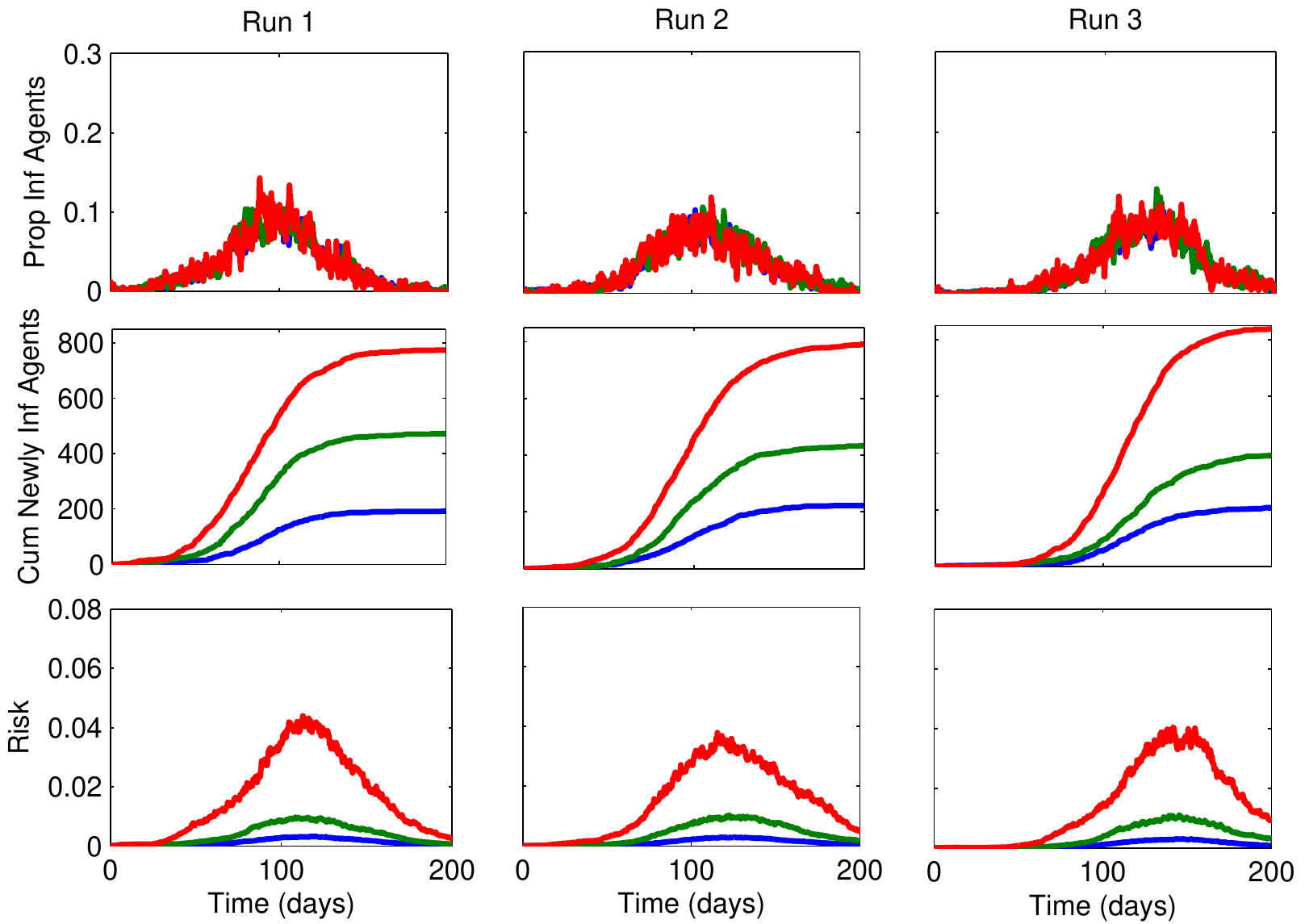}
  \end{center}
  \caption{ Three representative runs for the heterogeneous, high movement scenario with the cumulative number initially
    infected in each patch and the risk in each patch over time.}
  \label{F:High}
\end{figure}

\newpage

\begin{figure}[h]
  \begin{center}
     \includegraphics[scale=0.5]{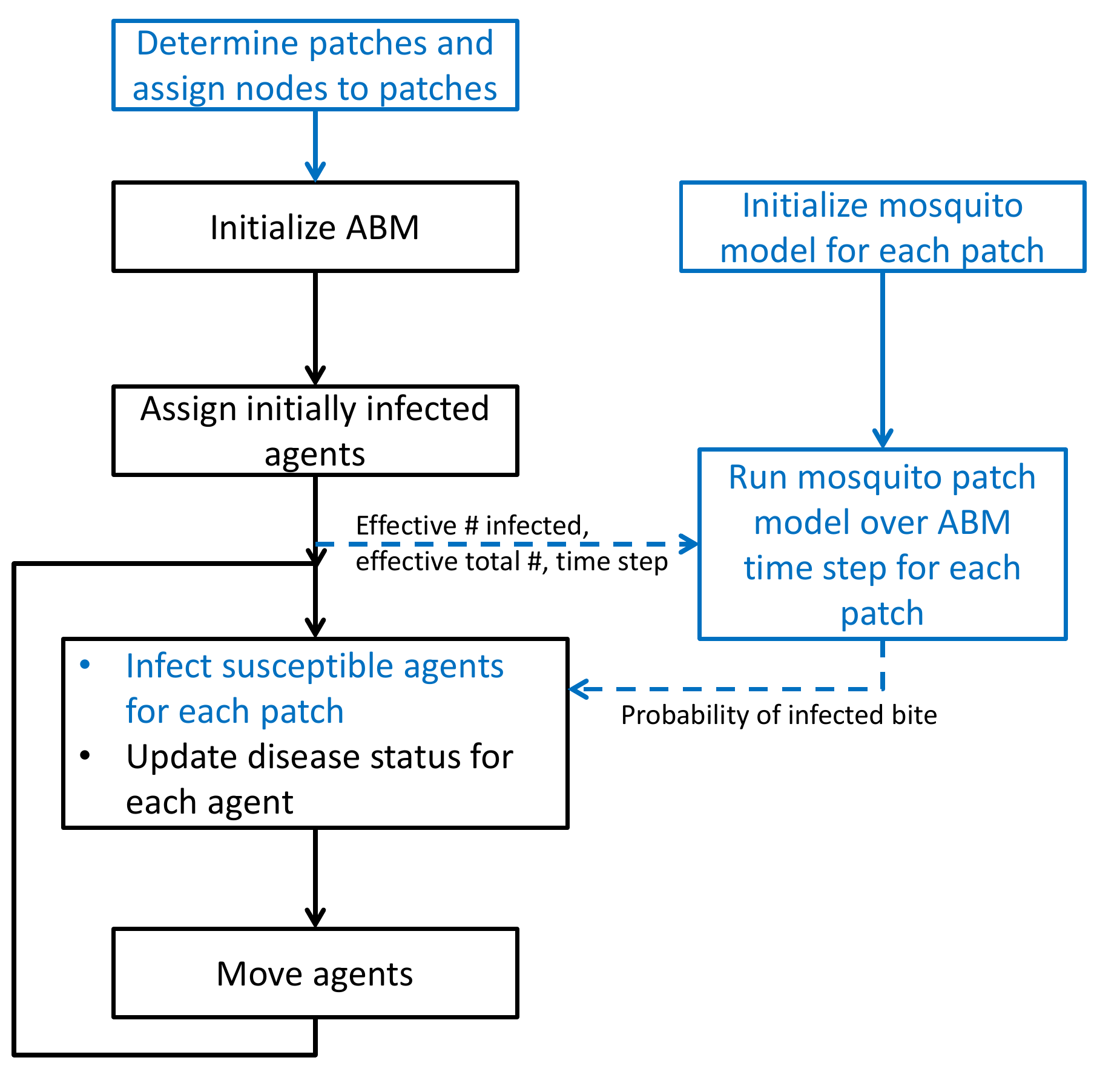}
  \end{center}
  \caption{ Flow chart for the ABM and associated patch model. Blue represents portions of the model that are added as a
    result of the network-patch approach. Dotted lines represent communication between the ABM and the patch model.}
  \label{F:flowchart}
\end{figure}

\label{lastpage}

\end{document}